\documentclass[twocolumn,twocolappendix]{aastex631}
\usepackage[utf8]{inputenc}
\usepackage{booktabs}
\usepackage{multirow}
\usepackage{float}

\begin{document}
\title{ALMA Spectral Survey of An eruptive Young star, V883 Ori (ASSAY): II. Freshly Sublimated Complex Organic Molecules (COMs) in the Keplerian Disk}

\correspondingauthor{Jeong-Eun Lee}
\email{lee.jeongeun@snu.ac.kr}

\author{Jae-Hong Jeong}
\affil{Department of Physics and Astronomy, Seoul National University, 1 Gwanak-ro, Gwanak-gu, Seoul 08826, Korea}

\author{Jeong-Eun Lee}
\affil{Department of Physics and Astronomy, Seoul National University, 1 Gwanak-ro, Gwanak-gu, Seoul 08826, Korea}
\affil{SNU Astronomy Research Center, Seoul National University, 1 Gwanak-ro, Gwanak-gu, Seoul 08826, Republic of Korea}

\author{Seonjae Lee}
\affil{Department of Physics and Astronomy, Seoul National University, 1 Gwanak-ro, Gwanak-gu, Seoul 08826, Korea} 

\author{Giseon Baek}
\affil{Department of Physics and Astronomy, Seoul National University, 1 Gwanak-ro, Gwanak-gu, Seoul 08826, Korea}
\affiliation{Research Institute of Basic Sciences, Seoul National University, Seoul 08826, Republic of Korea}

\author{Ji-Hyun Kang}
\affiliation{Korea Astronomy and Space Science Institute, 776 Daedeok-daero, Yuseong, Daejeon 34055, Korea}

\author{Seokho Lee}
\affiliation{Korea Astronomy and Space Science Institute, 776 Daedeok-daero, Yuseong, Daejeon 34055, Korea}

\author{Chul-Hwan Kim}
\affil{Department of Physics and Astronomy, Seoul National University, 1 Gwanak-ro, Gwanak-gu, Seoul 08826, Korea}

\author{Hyeong-Sik Yun}
\affiliation{Korea Astronomy and Space Science Institute, 776 Daedeok-daero, Yuseong, Daejeon 34055, Korea}

\author{Yuri Aikawa}
\affiliation{Department of Astronomy, University of Tokyo, 7-3-1 Hongo, Bunkyo-ku, Tokyo 113-0033, Japan}

\author{Gregory J. Herczeg}
\affiliation{Kavli Institute for Astronomy and Astrophysics, Peking University, Yiheyuan 5, Haidian Qu, 100871 Beijing, China}
\affiliation{Department of Astronomy, Peking University, Yiheyuan 5, Haidian Qu, 100871 Beijing, China}

\author{Doug Johnstone}
\affiliation{NRC Herzberg Astronomy and Astrophysics, 5071 West Saanich Road, Victoria, BC, V9E 2E7, Canada}
\affiliation{Department of Physics and Astronomy, University of Victoria, 3800 Finnerty Road, Elliot Building, Victoria, BC, V8P 5C2, Canada}

\author{Lucas Cieza}
\affiliation{Facultad de Ingenier\'ia y Ciencias, Instituto de Estudios Astrofísicos, Universidad Diego Portales, Av. Ejercito 441. Santiago, Chile}

\begin{abstract}
We present an investigation of Complex Organic Molecules (COMs) in the spatially resolved Keplerian disk around V883 Ori, an eruptive young star, based on a spectral survey carried out with ALMA in Band 6 (220.7$-$274.9 GHz). We identified about 3,700 molecular emission lines and discovered 23 COMs in the disk. We estimated the column densities of COMs detected through the iterative LTE line fitting method. According to our analyses, using only optically thin lines is critical to deriving the reliable column densities of COMs. Therefore, covering a large frequency range is important for the studies of COMs. The most distinct phenomenon found from the spectra of the V883 Ori disk is that nitrogen-bearing COMs other than CH$_{3}$CN are missing, whereas various oxygen-bearing COMs, except for the CH$_2$OH-bearing molecules, are detected. The missing CH$_2$OH-bearing COMs may indicate the warm water-ice dominant environment for forming COMs. 
We compared our results with various objects in different evolutionary stages, from Class 0 hot corinos to a Solar System comet 67P/Churyumov-Gerasimenko, to examine the effect of evolution on the COM compositions. In general, the COMs abundances relative to methanol in V883 Ori are higher than in the hot corinos and hot cores, while they are comparable to the cometary values. This may indicate the planet-forming material chemically evolves in the disk midplane after being accreted from the envelope. In addition, as found in the comet 67P/Churyumov-Gerasimenko, nitrogen might also be trapped as ammonium salt within the dust grains in the V883 Ori disk.
\end{abstract}

\section{Introduction}
Complex organic molecules (COMs) are prebiotic molecules with six or more atoms \citep{Herbst2009}. COMs can be detected via their emission lines when they are desorbed to the gas phase from icy mantles of dust grains. They have been detected in various protostellar evolutionary stages and environments across various star-forming regions: prestellar core \citep{Oberg_2010, Scibelli_2020}, hot corino \citep{PILS2, Belloche2020, Yang_2021, Hsu_2022, Lee_HOPS373}, hot core \citep{Belloche, Giseon2022, coccoa2023}, disk atmosphere \citep{Lee_2017, Lee_Chin_Fei2019}, disk midplane \citep{vantHoff_2018, jelee19, yamato2024chemistry}, and an extended shell around protostellar clusters \citep{McGuire2016}.\par

However, it remains elusive as to the degree that COM compositions would change from the initial compositions of the prestellar stage as the protostar evolves, as well as which chemistry (grain-surface versus gas-phase) would be dominant for COMs at different evolutionary stages. To answer these questions, we need to investigate COMs throughout the evolutionary stages. Although observational and theoretical studies have been done in various interstellar conditions to date \citep{Ceccarelli_PPVII}, the study of COMs associated with disks is difficult \citep{Aikawa2022arXiv221214529A, McGuire_2022}.\par

It is because the sublimation regions of COMs in the disks of typical low-mass protostellar systems, especially at disk midplanes, are confined to a few au of radii due to the disk temperature profile and density structure. Thus, even the Atacama Large Millimeter Array (ALMA) cannot resolve the COM sublimation regions in most disks with central protostars in the quiescent accretion phase.\par

Accretion outbursts can be used to study COMs compositions in a disk because the sublimation radii of COMs expand to large radii due to the luminosity increase by the outburst, and thus, freshly sublimated COMs emission can be detected (e.g., \citet{vantHoff_2018, jelee19, yamato2024chemistry, jelee2024}). From this idea, they could detect COMs emissions, revealing ice composition in the Keplerian disk of V883 Ori, including its midplane. However, these previous studies covered only narrow spectral ranges, limiting the full investigation of chemical characteristics in the V883 Ori disk.\par

An unbiased spectral line survey allows us to more fully understand a source's chemical properties. Excellent examples for ALMA spectral survey projects include Exploring Molecular Complexity with ALMA (EMoCA) for hot cores of Sgr B2(N) \citep{Belloche} and Protostellar Interferometric Line Survey (PILS) for a hot corino IRAS 16293-2422 \citep{PILS2}. These results have significantly contributed to our understanding of hot core and hot corino chemistry by discovering various new COMs from the unbiased continuous coverage of thousands of molecular lines \citep{jorgensen2020_araa}.\par

As an extension of these surveys, we carried out an unbiased spectral survey of V883 Ori to explore the disk chemical composition more thoroughly within the limit of observational sensitivity. V883 Ori is an eruptive young stellar object in transition between Class I and Class II, located within the Orion Molecular Cloud at a distance of 388 pc \citep{jelee19}. The bolometric luminosity of V883 Ori is still controversial, ranging from 186 L$_\odot$ \citep{Furlan2016} to 285 L$_\odot$ \citep{Strom1993, Sandell_2001} when they are adjusted to the distance of 388 pc. The outburst, which began before 1888, has lasted more than 130 years and may have been more luminous at its peak. \citep{Strom1993}. Its disk inclination is 38.3$^{\circ}$, and the mass of the protostar is 1.2 M$_{\odot}$ \citep{cieza_2016, jelee19}. Our spectral survey on V883 Ori fully covers 220.7 to 274.9 GHz using ALMA Band 6 with a uniform sensitivity rather than targeting specific species.

In this paper, we report the results of our COMs analyses with the V883 Ori spectral survey data, focusing on oxygen- and nitrogen-bearing species. In Section \ref{sec:observation and spectral extraction}, we describe our spectral survey observation and extraction of the spectra for the spectral analyses. In Section \ref{sec:spectral analyses}, we explain the method of line identification as well as how we dealt with attenuation by dust continuum. Then, a method of deriving column densities of the detected COMs is shown in Section \ref{sec:Iterative LTE line fitting}. The introduced method is especially useful for the wide range of line survey data. An effect of line optical depth on the column density estimation is also examined. In Section \ref{sec:results}, we present the results of our analyses: derived COMs compositions in the V883 Ori disk, and in Section \ref{sec:discussion}, we compare them with other objects and discuss the implications. Finally, we give a summary in Section \ref{sec:summary}.

\begin{figure*}[t]
    \centering
    \includegraphics[trim=0.2cm 0cm 0.2cm 0cm, clip=true,width=1.0\textwidth]{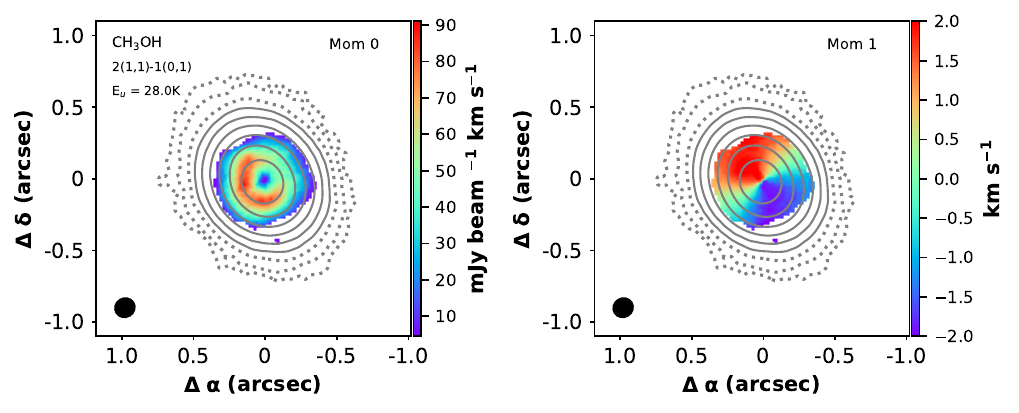}
    \caption{Integrated intensity (left panel) and intensity-weighted velocity (right panel) maps of the CH$_{3}$OH 2(1,1)-1(0,1) line, as a representative emission distribution for COMs in the V883 Ori disk. The structure in the left panel shows a ring-like shape, and the inner emission depression (r$<\sim$0.1$\arcsec$) is due to the optically thick dust blocking the molecular emission. The integral of the velocity ranges from -3.5 to 3.5 km s$^{-1}$. The images are clipped above the 4$\sigma$ (1$\sigma$=2.0 mJy beam$^{-1}$) level. Black contours represent continuum emission at 5, 10, 25, 50, 100, 200, 400, 800, and 1600$\sigma$ (1$\sigma$=0.023 mJy beam$^{-1}$). The three weakest black contours are shown as dotted lines, while the others are in solid lines. The beam is shown on the lower left corner of each panel. Please refer to \citet{jelee2024} (Paper I) for the analysis of the emission structure of various simple molecules as well as COMs. This figure is reproduced using the data presented in Paper I.}
    \label{fig:CH3OH map}
\end{figure*}

\section{Observation and Spectral Extraction} \label{sec:observation and spectral extraction}
In the ALMA Cycle 7 program 2019.1.00377.S (PI: Jeong-Eun Lee), we performed an unbiased spectral survey of V883 Ori in Band 6 with an angular resolution of 0.15$\arcsec$$-$0.2$\arcsec$, corresponding to $\sim$60 to 80 au at a distance of 388 pc \citep{jelee19}. The observations were conducted between May 2021 and November 2021. The on-source integration time is 194.5 minutes. The frequency range covered by the spectral survey is 220.703 GHz to 274.861 GHz with a spectral resolution of 488.281 kHz (0.66$-$0.53 km s$^{-1}$). This spectral range is about 30 times wider than the ALMMA Cycle 5 V883 Ori observation in Band 7 \citep{jelee19}. The continuum sensitivity is 0.01 mJy beam$^{-1}$, and the line sensitivity is 0.4$-$0.6 mJy beam$^{-1}$. The continuum emission was subtracted in the image space, and the continuum level is automatically measured using \texttt{STATCONT} \citep{Sanchez-Monge2018}. The dust continuum in Band 6 is less optically thick than in Band 7 by a factor of two, although the line intensity is two times stronger in Band 7. The details of observation and data reduction can be found in \citet[][Paper I hereafter]{jelee2024}. 

In V883 Ori, COMs emissions are mostly located inside the water emission region in the disk (\citealt{vantHoff_2018, jelee19, Tobin2023, yamato2024chemistry}, Paper I). Figure \ref{fig:CH3OH map} shows the emission distribution of a CH$_3$OH line, which represents the typical emission distribution of COMs in V883 Ori. The inner emission hole of COMs (r$<\sim$0.1$\arcsec$), as seen in the left panel, have been recognized in previous ALMA observations (e.g., Band 3: \citet{yamato2024chemistry}, Band 6: \citet{Tobin2023}; Paper I, and Band 7: \citet{vantHoff_2018,jelee19}). This structure is due to the extremely high dust opacity toward the disk center \citep{cieza_2016}.\par

We extracted averaged spectra over the entire COMs emission region in the disk ($\sim$0.3\arcsec) using a newly-developed filtering technique: first principal component (PC1) filtering method \citep{yun2023pca}. Using several strong, isolated COMs lines, \citet{yun2023pca} obtained a representative common geometric, kinematic feature of the COMs lines, a PC1, by utilizing the Principal Component Analysis (PCA) on the data cube. The PC1 was next used as a filter to apply to the 3D data cube, resulting in a PC1-filtered spectrum. This method allowed us to correct kinematic features in the disk spectrum without detailed disk kinematic modeling and, therefore, to achieve a higher signal-to-noise ratio (SNR) than a spectrum produced by the velocity alignment method \citep{jelee19}. For more information, please refer to \citet{yun2023pca}.\par

\begin{figure*}[t]
    \centering
    \includegraphics[trim=0.2cm 0cm 0.2cm 0cm, clip=true,width=1.0\textwidth]{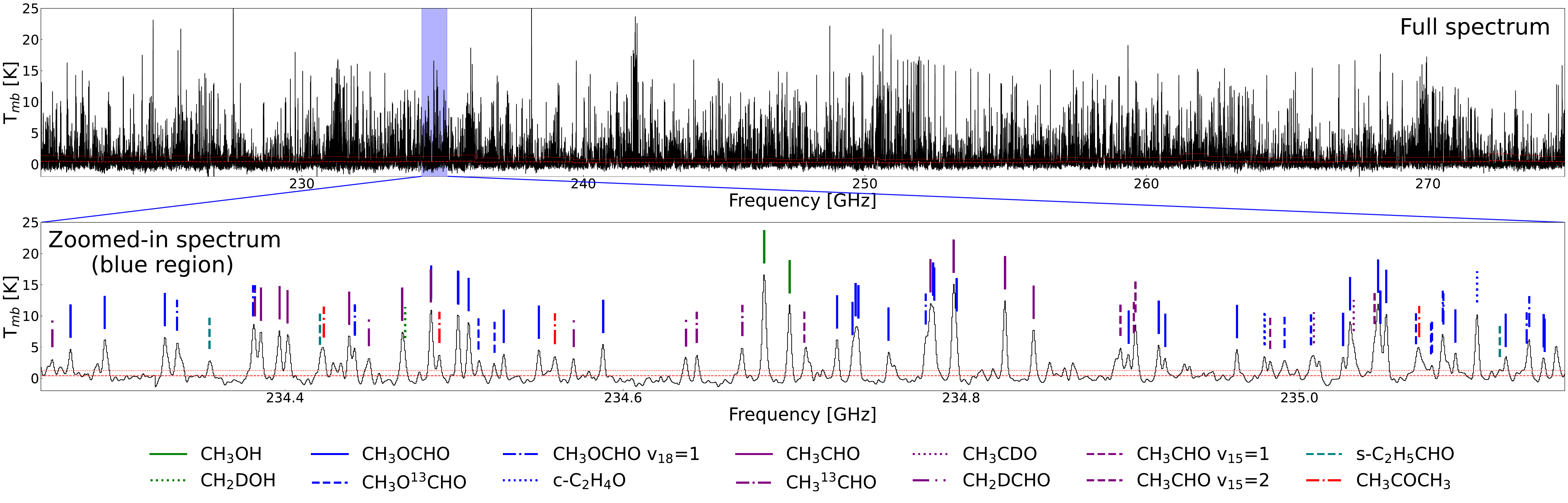}
    \caption{The disk-averaged spectrum of V883 Ori. This is the observed spectrum extracted from the PC1-filtering, but not the reconstructed one through the dust attenuation and beam-filling factor correction. The upper panel shows the spectrum of the full frequency range covered in our observation of V883 Ori, after subtraction of the continuum level. The lower panel shows a randomly selected, zoomed-in spectrum of the blue-shaded region in the upper panel. The vertical color bars represent identified lines of COMs from our study. The red dashed/dotted horizontal lines represent the 1$\sigma$/3$\sigma$ levels of the observed spectrum.}
    \label{fig:spectrum_before}
\end{figure*}

The resulting full PC1-filtered spectrum is shown in the upper panel of Figure \ref{fig:spectrum_before}, and a zoomed-in portion is provided in the lower panel. The spectrum is crowded but contains well-resolved (the line width is $\sim$3 km s$^{-1}$) rotational transition lines of molecules in the vibrational ground or excited levels.

\section{SPECTRAL ANALYSES} \label{sec:spectral analyses}
\subsection{Dust attenuation of line intensities}
As seen by \citet{jelee19}, the dust opacity affects the line intensity significantly in V883 Ori. Therefore, after the spectral extraction, we corrected the effect of dust attenuation on the intensity of the molecular lines. The observed line intensities unaffected by dust attenuation ($T_{\rm line}$) can be obtained by dividing the observed continuum-subtracted intensities ($T_{\rm obs} - T_{\rm cont}$) by $e^{-\tau_{\rm \nu,dust}}$ (Equation \ref{eq:dust correction1}), where $T_{\rm cont}\equiv \eta\times T_{\rm rot}\times(1-e^{-\tau_{\rm \nu, dust}})$. 
Here, $T_{\rm obs}$ is the total observed intensity, $T_{\rm cont}$ is the observed continuum intensity, $T_{\rm rot}$ is the source function, and $\eta$ is the beam-filling factor. We assume that dust and gas are well coupled and that the gas is in LTE. Therefore, $T_{\rm line}$ can be calculated by
    \begin{equation}
    \label{eq:dust correction1}
        T_{\rm line} = \eta\times T_{\rm rot}\times(1-e^{-\tau_{\rm \nu,line}}) = {{T_{\rm obs} - T_{\rm cont}}\over{e^{-\tau_{\rm \nu,dust}}}}.
    \end{equation}

\begin{figure}[t]
    \centering
    \includegraphics[trim=0.2cm 0cm 0.2cm 0cm, clip=true,width=0.48\textwidth]{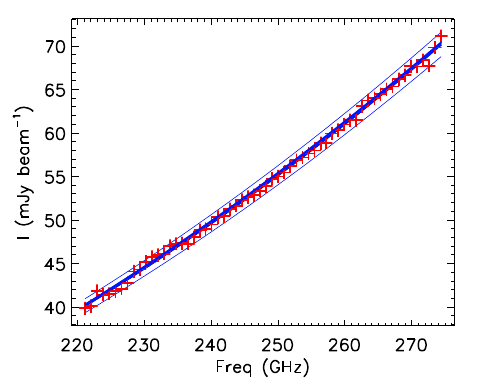}
    \caption{The continuum intensities as a function of spectral window weighted by the moment 0 map of the PCA Eigen cube 1. The red crosses indicate the corrected intensity of each spectral window. The thick blue line is the best-fit result of $T_{\rm cont}\equiv \eta\times T_{\rm rot}\times(1-e^{-\tau_{\rm \nu, dust}})$. The thin blue lines represent $\pm$2$\%$ range.}
    \label{fig:dust fit}
\end{figure}

Calculating $\tau_{\rm \nu,dust}$ requires the molecular hydrogen column density ($N_{\rm H_{2}}$),  the dust mass opacity for a certain type of dust ($\kappa_{\rm ref}$), and the spectral index of $\kappa_{\rm ref}$ ($\beta$), as detailed here
(Equation \ref{eq:dust correction2}). 
    \begin{equation}
    \label{eq:dust correction2}
        \tau_{\rm \rm \nu,dust}=(N_{\rm H_{2}}\cdot\kappa_{\rm ref}\cdot m_{\rm H_{2}}\cdot0.01)\cdot({\nu\over\nu_{\rm ref}})^{\beta}.
    \end{equation}
We adopt $\kappa_{\rm ref}= 2.2$ cm$^{2}$ g$^{-1}$ (at $\nu_{\rm ref}$= 230 GHz) and $\beta$=1 from \citet{Cieza2018}. To get $N_{\rm H_{2}}$ and $\eta$, we fitted the continuum intensities as a function of spectral windows, weighted by the moment 0 map of the PC1 (Figure \ref{fig:dust fit}). The moment 0 map of the PC1 represents the region where the most COM emission originates \citep{yun2023pca}. To correct a systematic difference among spectra of three science goals (SGs) seen in the original reduction (Paper I), we scaled up the SG2 and SG3 spectra by 5\% and 2\%, respectively. Finally, the fitted $\eta$ is 0.384, and the fitted $N_{\rm H_{2}}$ is $1.24\times10^{25}$ cm$^{-2}$ with uncertainty of 2$\%$. Although $\tau_{\rm \nu,dust}$ can vary with radius, our fitted $\tau_{\rm \nu,dust}$ represents the averaged $\tau_{\rm \nu,dust}$ over the COM emission region ($\sim$0.1$\arcsec$ to $\sim$0.3$\arcsec$; Paper I), and it is appropriate for the dust attenuation correction of the PC1-filtered spectra. Also, there is no notable size variation of the optically thick region ($\sim$0.1$\arcsec$) across our frequency range.\par

According to the calculation, the observed line intensities ($T_{\rm obs} - T_{\rm cont}$, Figure \ref{fig:spectrum_before}) are reduced to 34$\%$ $-$ 42$\%$ of the dust-corrected values ($T_{\rm line}$) over the observed frequency range. We used these corrected spectra throughout our analyses with molecular line identification and iterative LTE line fitting. However, the spectra shown in the figures in this paper (Figure \ref{fig:spectrum_before}, \ref{fig:line identification} and \ref{fig:fitted_spectra_comparison}) are presented using the originally observed PC1-filtered spectra ($T_{\rm obs} - T_{\rm cont}$), not the reconstructed ones.

\subsection{Molecular line identification}\label{sec:molecular line identification}
To investigate the chemical composition of the disk, the first step is to identify which molecules produce the observed lines. We used the eXtended CASA Line Analysis Software Suite \citep[\texttt{XCLASS}, ][]{Moller17} for the line identification. Assuming local thermodynamic equilibrium (LTE), \texttt{XCLASS} can model the spectra of each molecular species by managing several parameters, such as velocity width ($v_{\rm width}$), rotation temperature ($T_{\rm rot}$), column density ($N$), and beam-filling factor ($\eta$).\par 

To simplify the analysis, $N$ was used as the only free parameter. We fixed $v_{\rm width}$  at 3 km s$^{-1}$, which is obtained from the PC1-filtered spectrum. $T_{\rm rot}$ was set to 120 K, which is the average temperature within 0.3$\arcsec$ radius of the disk model of V883 Ori \citep{jelee19}. The radius of 0.3$\arcsec$ corresponds to the representative COMs emission size (moment 0 map of PC1) in V883 Ori \citep{yun2023pca}. We fixed $\eta$ at 0.384, which was estimated from the continuum fitting after weighting with the moment 0 map of PC1 (Figure \ref{fig:dust fit}). 
This value is consistent with the one derived from the optically thick, saturated CH$_3$OH lines, as the strongest CH$_3$OH line shows the closest correlation with PC1 \citep{yun2023pca}. After correcting for dust attenuation, the intensities of the optically thick CH$_{3}$OH lines are $\sim$46 K, which corresponds to 0.384 times the source function (120 K).\par

\texttt{XCLASS} uses the molecular databases from both Cologne Database for Molecular Spectroscopy (CDMS) \citep{Muller01, MULLER05} and Jet Propulsion Laboratory (JPL) \citep{PICKETT1998883}. CDMS and JPL databases contain about 1,500 molecules (ranging from simple molecules to COMs) that emit at least one molecular line in our observed spectral windows.\par

Due to the densely populated energy states of COMs compared to the simple molecules, COMs can emit numerous transition lines across the frequency range of this observation. In addition, with 1,500 species to consider, multiple candidate molecules may be assigned for a single emission line, which makes the line identification more complicated.
Therefore, to properly identify the observed lines, we created a model spectrum with the \texttt{myXCLASS} function within \texttt{XCLASS} for individual test species. Then, we checked whether the frequencies of all modeled transitions of the test species matched the frequencies of the observed lines within the spectral resolution.\par

\begin{figure*}[t] 
    \centering
    \includegraphics[trim=0.2cm 0cm 0.2cm 0cm, clip=true,width=0.325\textwidth]{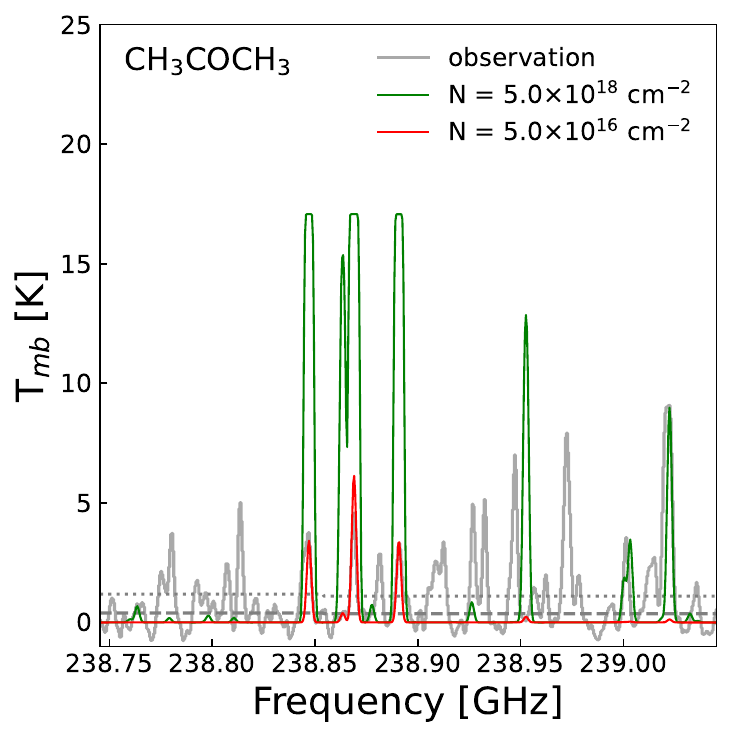} 
    \includegraphics[trim=0.2cm 0cm 0.2cm 0cm, clip=true,width=0.325\textwidth]{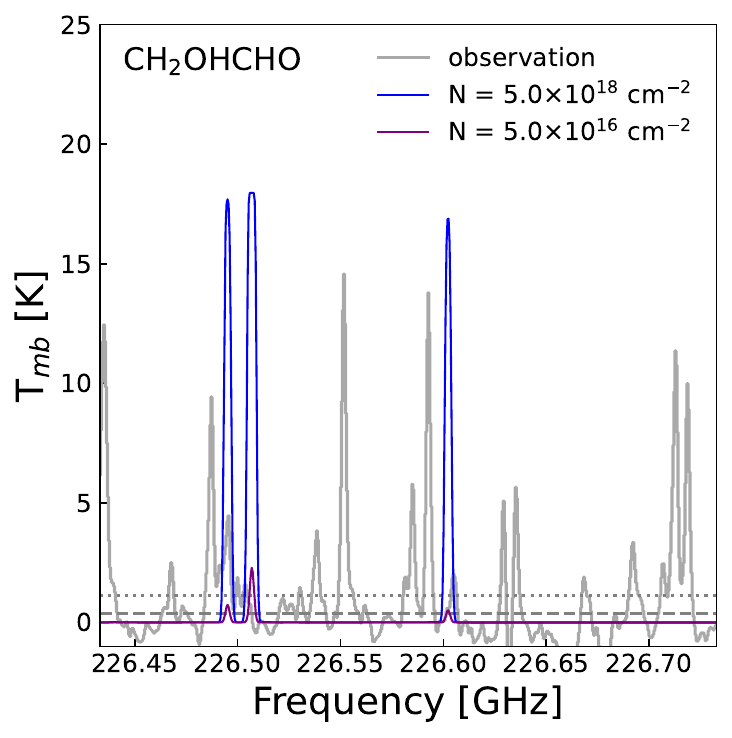} 
    \includegraphics[trim=0.2cm 0cm 0.2cm 0cm, clip=true,width=0.325\textwidth]{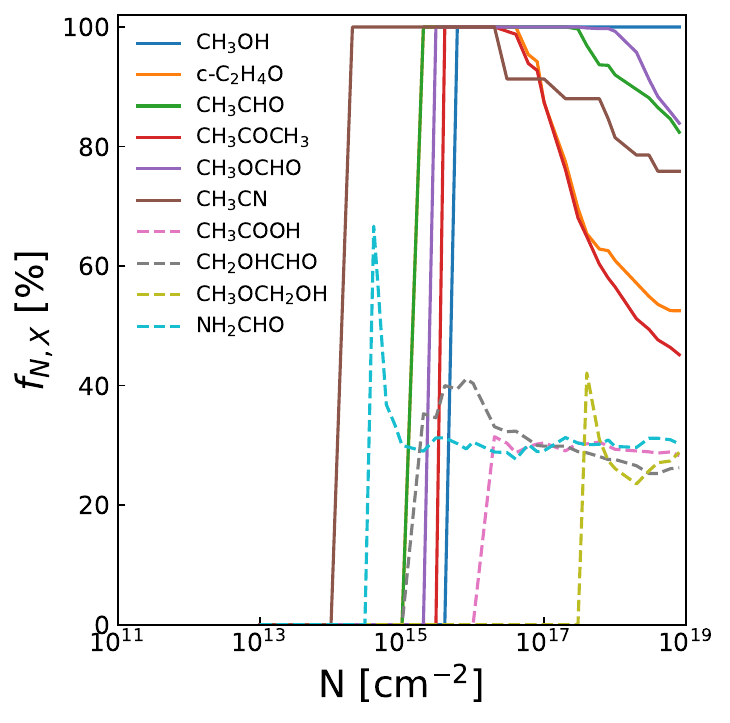}
    \caption{Figures to depict the line identification method. Left and middle panels: Four examples for XCLASS model spectra, overlaid on the observed spectra. The colored solid lines indicate acetone (CH$_{3}$COCH$_{3}$) and glycolaldehyde (CH$_{2}$OHCHO) model spectra at two different column densities (N=5.0$\times$10$^{18}$ cm$^{-2}$, 5.0$\times$10$^{16}$ cm$^{-2}$), respectively. The gray solid lines are the observed spectra and gray dashed/dotted horizontal lines represent the 1$\sigma$/3$\sigma$ levels of the observed spectrum. Right panel: As examples for $f_{\rm N, X}$ (Equation \ref{eq:ratio}), we show the $f_{\rm N, X}$ profiles for 10 molecules. The profiles are shown as solid lines for the six molecules whose ratios reached 100$\%$ (considered as detected) at some column density steps, and as dashed lines for the remaining four molecules.}
    \label{fig:line identification}
\end{figure*}

The left and middle panels of Figure \ref{fig:line identification} show examples of model spectra compared to observed spectra. Among the four cases (model spectra in green, red, blue, and purple) in the figure, only the model for acetone (CH$_{3}$COCH$_{3}$), modeled with an input column density of 5$\times$10$^{16}$ cm$^{-2}$ (red solid lines in the left panel), matches the observation. On the other hand, the acetone model with a column density of 5$\times$10$^{18}$ cm$^{-2}$ (green solid line in the left panel) only provides a partial match. In this case, it is possible to make a misidentification of an observed line as acetone, such as a line near 239.02 GHz. Lastly, in the case of glycolaldehyde (CH$_{2}$OHCHO) in the middle panel, neither of the two cases (blue and purple solid lines) can successfully identify the observed lines. Therefore, it is important to use both the correct species and the appropriate column density for the line identification, as \citet{jorgensen2020_araa} pointed out.\par
    
Thus, for every possible species (about 1,500), we constructed model spectra with 35 column density steps (from 5$\times$10$^{12}$ cm$^{-2}$ to 5$\times$10$^{18}$ cm$^{-2}$). Then, we defined a simple parameter $f_{\rm N, X}$ (Equation \ref{eq:ratio}) and calculated it for all model spectra at each column density ($N$) step and species ($X$), so that we could readily filter for strong candidate molecules.
    \begin{equation}
    \label{eq:ratio}
        f_{\rm N, X}\equiv{\rm\#\:of\:observed\:lines\:matched\:with\:model \over \#\:of\:modeled\:lines}
    \end{equation} 
For example, in the frequency range corresponding to the x-axis of the left panel of Figure \ref{fig:line identification}, the value of $f_{\rm N, X}$ for acetone would be 3/3 (100\%) at N=5$\times$10$^{16}$ cm$^{-2}$ and 5/8 (62\%) at N=5$\times$10$^{18}$ cm$^{-2}$. Thus, the former model spectrum can be adopted as the acetone line identification.\par

The right panel of Figure \ref{fig:line identification} shows the $f_{\rm N, X}$ profiles for 10 test molecules {\it over the entire frequency range} of the spectral survey. These $f_{\rm N, X}$ were automatically calculated using the lines regardless of line intensity if the intensities are above the 3$\sigma$ levels. From the $f_{N,X}$ profiles in the figure, we can consider CH$_{3}$OH, c-C$_{2}$H$_{4}$O, CH$_{3}$CHO, CH$_{3}$COCH$_{3}$, CH$_{3}$OCHO, and CH$_{3}$CN as strong candidates for molecules present in V883 Ori, whereas CH$_{3}$COOH, CH$_{2}$OHCHO, CH$_{3}$OCH$_{2}$OH, and NH$_{2}$CHO can be ruled out as non-detections.\par

We, therefore, collected molecules that reached 100$\%$ of $f_{\rm N, X}$ at some column density steps among the 1,500 species. Usefully, $f_{N,X}$ decreases after reaching 100$\%$ because the model lines with higher column densities overproduce non-detected weak lines, as illustrated by the CH$_{3}$COCH$_{3}$ model with 5$\times$10$^{18}$ cm$^{-2}$ (the green spectra in the left panel). Finally, we performed a visual inspection to confirm the detection of those molecules selected by this procedure. Furthermore, for the species with relatively high $f_{\rm N, X}$ still below 100$\%$, we retested using finer column density steps and detected a few more species.\par
  
\section{The LTE analyses of COMs}\label{sec:Iterative LTE line fitting}
\subsection{Column density estimation with an iterative LTE line fitting}
For the detected molecules, we derive the average column densities over the COMs-emission region in the disk of V883 Ori using an LTE line fitting tool. In this section, we describe the method, and the results are presented in Section \ref{sec:results}.\par

We fitted the observed lines with the model spectra of each detected species individually by adjusting the input parameters for the model spectra; the results are independent of other species because we use only isolated and optically thin lines, excluding the blended lines in our iterative line fitting process. As mentioned in the previous section, the column density ($N$) was the only free parameter in the analysis, while the other parameters were fixed to the aforementioned values. We operated a fitting tool of \texttt{XCLASS}: the Modeling and Analysis Generic Interface for eXternal numerical codes (\texttt{MAGIX}) \citep{Moller13}. The genetic algorithm was used to derive the best-fit values of the parameters, and the Markov chain Monte Carlo (MCMC) method was used to calculate the uncertainties. These algorithms are available as part of \texttt{MAGIX}.\par

The spectra were extracted over the COMs emission region, which is the dense inner disk. Therefore, the spectra contain optically thick lines, which must be excluded in the fitting process to accurately estimate column densities (e.g., \citet{PILS14, PILS24}). However, since we do not know which lines are optically thick in the first place, we repeated the line fitting, removing lines with an optical depth higher than a certain value and, thus, utilizing only the optically thin lines at each iteration.\par
    
\begin{figure}[t]
    \centering
    \includegraphics[trim=0.1cm -0.4cm 0cm 0cm, clip=true,width=0.49\textwidth]{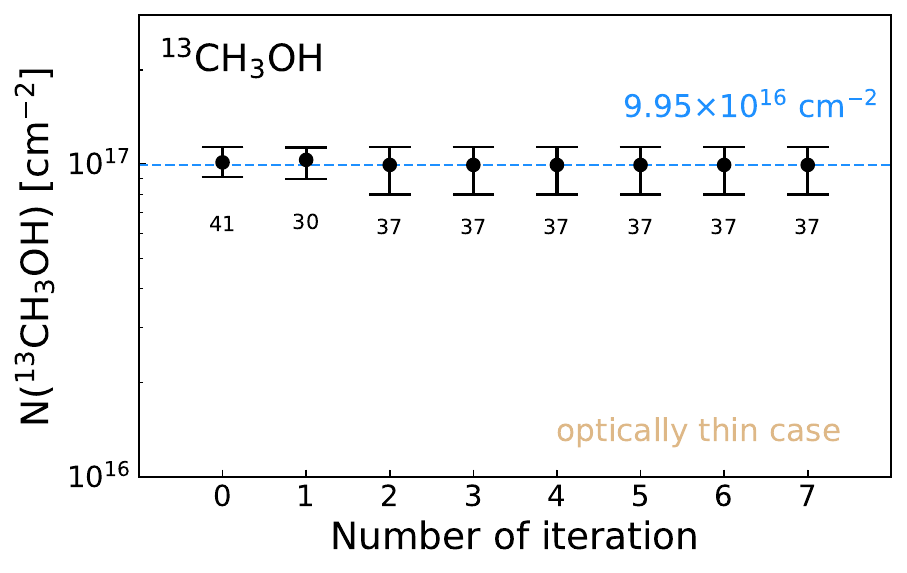}
    \includegraphics[trim=0.1cm -0.4cm 0cm 0cm, clip=true,width=0.49\textwidth]{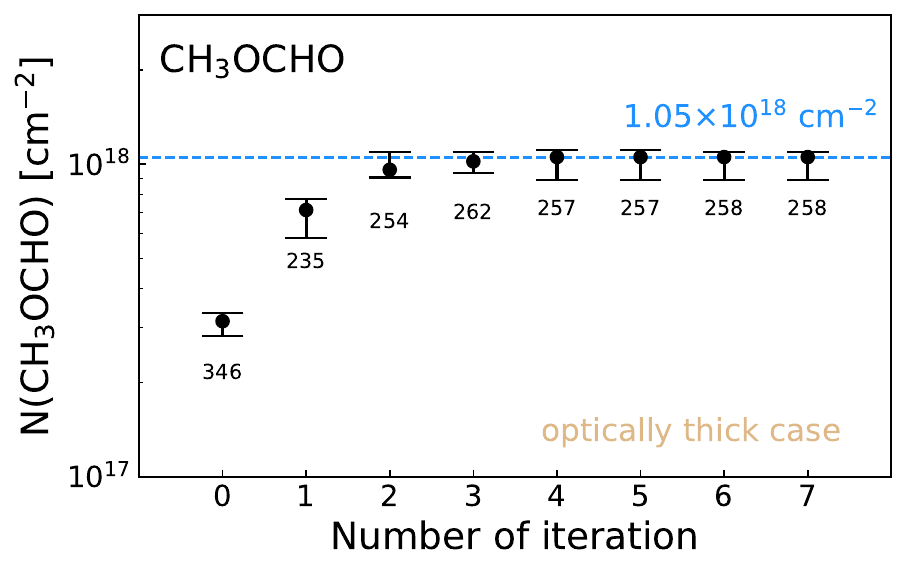}
    \caption{Examples of iterative LTE line fitting results on a molecule that emits mostly optically thin lines ($^{13}$CH$_{3}$OH), and several optically thick lines (CH$_{3}$OCHO). The 0th iteration indicates the fitting result when using the isolated lines without optical depth consideration. The remaining iterations show the results of the fitting using optically thin ($\tau_{\rm \nu,line}<0.7$), isolated lines extracted from the results of the previous iteration. The numbers under data points are the number of isolated molecular lines used in the LTE fitting in that iteration. The number in the upper right of each figure is the column density obtained as the final result of this method, which is also drawn as a blue horizontal dashed line.}
    \label{fig:LTE fitting examples}
\end{figure}
    
Here, we present a simple algorithm for this iterative line-fitting method. First, we fitted an LTE model spectrum of each molecular species over the full observed spectra for all detected molecules, individually, regardless of line optical depths at the line center ($\tau_{\rm \nu,line}$). Second, we calculated the line optical depths from the first fitted model spectrum for each species. We then collected observed lines, which are optically thin and well isolated in the model. In addition, we removed observed lines with large line width ($>$ 4 km s$^{-1}$), which may be due to blending with unidentified lines. These collected lines of each molecule were used for the next fitting process. Then again, isolated, optically thin lines were re-collected from these new results. This fitting process was done iteratively until the best-fit value of the column densities converged. This approach using only optically thin, isolated lines is possible because we identified as many molecular lines as possible during the line identification process, in addition to the broad frequency coverage of our spectral line survey.\par

The optical depth criterion is set to $\tau_{\rm \nu,line}=0.7$, meaning that only lines with their modeled $\tau_{\rm \nu,line}$ lower than 0.7 are considered in the fitting process. We adopt 0.7 because lowering the criterion below 0.7 reduces the number of lines available for fitting, and the selected lines have overall lower signal-to-noise ratios. For example, if we take the optical depth criterion as 0.2, the signal-to-noise ratios of the lines are mostly in the range of 3$-$5, and the number of lines used for fitting is reduced by about a factor of two, compared to the case of 0.7.\par

Two examples of column densities obtained from the iterative fitting method are presented in Figure \ref{fig:LTE fitting examples}. These two molecules ($^{13}$CH$_{3}$OH and CH$_{3}$OCHO) are exemplary of the molecules emitting mostly optically thin lines and the molecules emitting many optically thick lines, respectively. The column density of the former case ($^{13}$CH$_{3}$OH) remains nearly constant through the iteration because only few optically thick lines, out of 41 isolated lines, are excluded from the fit, i.e., most of the lines are optically thin.\par

On the other hand, the column density of the latter case (CH$_{3}$OCHO) gradually increases from 3.15$\times$10$^{17}$ cm$^{-2}$ and converges to 1.05$\times$10$^{18}$ cm$^{-2}$, which is about three times larger than the result of 0th iteration. The number of isolated and optically thin lines of CH$_{3}$OCHO increases as the column density increases because unseen lines in the model with smaller column densities become more prominent as the column density increases. The fitted column density increases as new optically thin lines are included and new optically thick lines are excluded at the next iteration. The number of used isolated lines decreases from 346 (in iteration 0) to 235 (in iteration 1) because iteration 0 uses isolated lines from all $\tau_{\rm \nu,line}$ ranges. In contrast, iteration 1 uses isolated lines with $\tau_{\rm \nu,line}<0.7$. Also, some lines turn out to be blended, which reduces the number of isolated lines itself. This is due to newly populated lines from other molecules whose column density is increased.\par

Lastly, for the molecules that were not detected in V883 Ori, we obtained upper limits on their column densities to compare with previous studies on COMs. The column density producing the strongest line corresponding to the 3$\sigma$ level of the observation is considered the upper limit.\par

To validate our estimation of molecular column densities via iterative LTE line fitting, we also utilized a rotation diagram method, a simpler LTE analysis. We described the results and implications from the rotation diagram analysis in Appendix \ref{sec:rotation diagram}. In short, the difference in column densities obtained from the two methods has a scatter of 0.28 dex with only a slight offset of 0.03 dex.

\begin{figure}[t]
    \centering
    \includegraphics[trim=0.2cm 0cm 0.2cm 0cm, clip=true,width=0.48\textwidth]{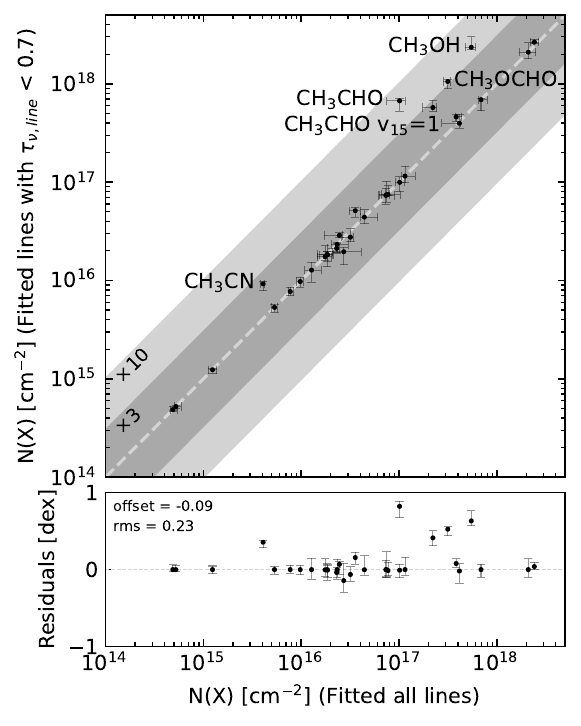}
    \caption{Comparison of the column densities of molecules obtained by iterative LTE line fitting using only isolated, optically thin lines ($\tau_{\rm \nu,line}<0.7$, corresponds to the column density result of the 7th iteration of Figure \ref{fig:LTE fitting examples}) and using isolated lines of all optical depths (corresponds to the result of the 0th iteration of Figure \ref{fig:LTE fitting examples}). Each data point represents the derived column densities of a single species. Labeled molecules are the ones that show differences of larger than 0.2 dex. The line representing 1:1 is indicated by a gray dashed line. The lines representing 3 times and 10 times differences are shown as gray solid lines with dark and light gray shaded areas, respectively.}
    \label{fig:comparison_different_tau}
\end{figure}

\subsection{Line optical depth effects on column density estimation}\label{sec:optical depth effect}
In this section, we quantitatively investigate the effect of the line optical depth in estimating column density by comparing two cases (1) only with optically thin lines and (2) with all lines.\par

Figure \ref{fig:comparison_different_tau} compares the estimated column density results for the two cases. Most molecules are aligned along a one-to-one line, so the resulting column densities of the two approaches are likely to be consistent. However, for some molecules such as methanol (CH$_{3}$OH), methyl formate (CH$_{3}$OCHO), and acetaldehyde (CH$_{3}$CHO), the latter method underestimates the column density by up to almost an order of magnitude compared to the former. These are the molecules whose column densities tend to increase as the iterations continue and converge to a particular value during the iterative fitting process, as shown in the example on the lower panel of Figure \ref{fig:LTE fitting examples}. Note that even vibrationally excited molecules like CH$_{3}$CHO v$_{15}$=1 can emit optically thick lines, and their column density can be therefore underestimated. In Appendix \ref{sec:effect of line optical depth seen with spectra}, we examine in more detail the optical depth effect in the modeling of observed spectra.
\par
    
\begin{figure}[t]
    \centering
    \includegraphics[trim=0.2cm 0cm 0.2cm 0cm, clip=true,width=0.48\textwidth]{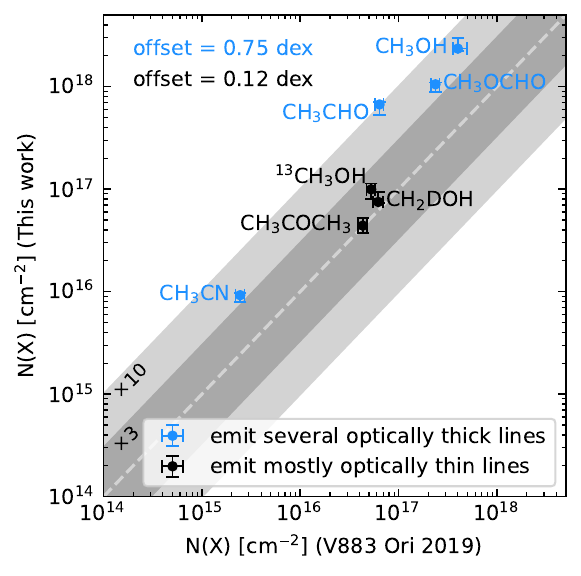}
    \caption{The estimated column density results from 2019 V883 Ori \citep{jelee19} compared to those derived in this work (Table \ref{tb:identified}, y-axis of Figure \ref{fig:comparison_different_tau}). The data points are from the Supplementary Table 1 by \citet{jelee19}. The gray dashed line represents the one-to-one line. The lines representing 3 times and 10 times difference from the one-to-one values are shown as solid gray lines with dark and light gray shaded areas, respectively.}
    \label{fig:comparison_2019,2022}
\end{figure}

Furthermore, we compared the column densities from this work to the column densities by \citet{jelee19} that used the ALMA Band 7 spectra of V883 Ori (see Figure \ref{fig:comparison_2019,2022}). Previously, due to the small frequency range of the observation, it was impossible to do selective line fitting using only isolated, optically thin lines. Therefore, Figure \ref{fig:comparison_2019,2022} shows that the values from \citet{jelee19} follow the same tendency as seen in Figure \ref{fig:comparison_different_tau}. The column densities of molecules emitting optically thick lines are underestimated by an average difference of 0.75 dex, or 5.6 times, while molecules emitting mostly optically thin lines have similar best-fit column densities with an average difference of 0.12 dex, compared to the values derived from only optically thin lines by this work. For example, the derived column density of CH$_{3}$OH is 5.9 times higher than that in \citet{jelee19}.\par

The above comparisons demonstrate that the wide frequency coverage obtained by this spectral survey is critical for accurate column density estimation with a sufficient number of optically thin lines. Moreover, the iterative LTE line fitting method allows us to effectively exclude optically thick lines that can lead to underestimating the column densities. 
\par

\section{Results}\label{sec:results}
In this paper, we present the analysis results of only COMs, although we have identified lines from simple molecules as well as COMs to extract isolated lines properly in the procedures of line identification and column density estimation described in previous sections.

\subsection{COMs detected in the disk of V883 Ori}\label{sec:line identification result}
As a result of the line identification, a total of 23 complex organic molecular species have been found in V883 Ori, based on about 3,700 lines above 3$\sigma$. Most observed lines have been identified, but $\sim$20$\%$ of the lines remain unidentified. Many of these unidentified lines lack corresponding line information in the databases we adopted. Therefore, updated spectroscopic databases are necessary to identify the remaining lines. The spectroscopic data from direct spectroscopic experiments (e.g., \citealt{Ohno_2022, Müller2022, Ferrer2023, Oyama_2023}) could be adopted, but we leave this as a future work.\par
    
Table \ref{tb:identified} lists all detected COMs and their column densities obtained through the iterative LTE line fitting method. The 23 species consist of 11 main isotopologues, with the remaining 12 composed of various $^{13}$C, $^{18}$O isotopologues, and deuterated molecules.\par

This plethora of identified COMs demonstrates that the disk of V883 Ori is as chemically rich as the Class 0 hot cores \citep{Belloche, Giseon2022, coccoa2023} and hot corinos \citep{PILS19, Hsu_2022}. However, for V883 Ori oxygen-bearing COMs were exclusively detected, especially molecules containing one oxygen. Methyl formate (CH$_{3}$OCHO) is the only detected molecule that contains two oxygen atoms. On the other hand, although numerous detectable lines of nitrogen-bearing COMs are covered in the frequency range, only methyl cyanide (CH$_3$CN) was detected in this disk.\par 

\startlongtable
\begin{deluxetable*}{cccccccccc}
    \tabletypesize{\scriptsize}
    \tablecaption{List of all COMs detected in V883 Ori with minimum line information and column densities from the LTE line fitting analysis.
    \label{tb:identified}}
    \tablehead{
    \colhead{No.} & \colhead{Species} &  \colhead{$\sharp$ of lines\tablenotemark{a}}& \colhead{E$_{u}$ (K)}&  \colhead{N (cm$^{-2}$)} & \colhead{No.} & \colhead{Species} &  \colhead{$\sharp$ of lines}& \colhead{E$_{u}$ (K)}&  \colhead{N (cm$^{-2}$)}}
    \startdata
         \multicolumn{5}{c}{Methanol} &   & CH$_{3}$$^{13}$CHO v$_{15}$=1 & 27 (12) & 274.8 - 366.3 & $1.96_{-0.51}^{+0.21}\times10^{16}$\\
 	1 & CH$_{3}$OH & 102 (20) & 20.1 - 926.6 & $2.35_{-0.14}^{+0.65}\times10^{18}$ & 12 & CH$_{3}$CDO & 86 (41) & 74.5 - 233.4 & $9.75_{-1.25}^{+1.03}\times10^{15}$\\ 
      & CH$_{3}$OH v$_{12}$=1 & 25 (13) & 326.2 - 771.0 & $2.65_{-0.10}^{+0.17}\times10^{18}$ & 13 & CH$_{2}$DCHO & 126 (67) & 73.0 - 294.3 & $2.12_{-0.20}^{+0.27}\times10^{16}$\\ 
 	  & CH$_{3}$OH v$_{12}$=2 & 9 (1) & 542.9 - 717.5 & $2.09_{-0.30}^{+0.53}\times10^{18}$ & \multicolumn{5}{c}{Ethanol}\\ 
 	2 & $^{13}$CH$_{3}$OH & 68 (37) & 19.9 - 480.3 & $9.95_{-1.95}^{+1.43}\times10^{16}$ & 14 & C$_{2}$H$_{5}$OH & 51 (37) & 28.5 - 216.4 & $2.75_{-0.28}^{+0.61}\times10^{16}$\\ 
 	3 & CH$_{3}$$^{18}$OH & 29 (16) & 33.4 - 196.9 & $7.33_{-1.38}^{+1.30}\times10^{16}$ & \multicolumn{5}{c}{Propanal}\\ 
 	4 & CH$_{2}$DOH & 87 (47) & 22.6 - 347.1 & $7.50_{-0.21}^{+1.81}\times10^{16}$ & 15 & s-C$_{2}$H$_{5}$CHO & 206 (87) & 41.0 - 302.3 & $2.87_{-0.15}^{+0.19}\times10^{16}$\\ 
 	     \multicolumn{5}{c}{Methyl formate} & \multicolumn{5}{c}{Propyne}\\ 
    5 & CH$_{3}$OCHO & 676 (258) & 36.3 - 1391.0 & $1.05_{-0.17}^{+0.04}\times10^{18}$ & 16 & CH$_{3}$CCH & 16 (8) & 74.6 - 176.6 & $1.27_{-0.31}^{+0.26}\times10^{16}$\\
 	  & CH$_{3}$OCHO v$_{18}$=1 & 265 (198) & 220.2 - 740.9 & $6.89_{-1.56}^{+0.20}\times10^{17}$ & \multicolumn{5}{c}{Dimethyl ether}\\ 
 	6 & CH$_{3}$O$^{13}$CHO & 177 (70) & 36.0 - 295.9 & $7.42_{-1.16}^{+2.83}\times10^{16}$ & 17 & CH$_{3}$OCH$_{3}$ & 292 (32) & 21.8 - 1882.4 & $4.60_{-0.48}^{+0.28}\times10^{17}$\\ 
 	     \multicolumn{5}{c}{Ethylene oxide} & \multicolumn{5}{c}{Acetone}\\
    7 & c-C$_{2}$H$_{4}$O & 40 (23) & 31.5 - 495.4 & $5.12_{-0.74}^{+0.45}\times10^{16}$ & 18 & CH$_{3}$COCH$_{3}$ & 187 (108) & 34.2 - 997.3 & $4.40_{-0.65}^{+0.86}\times10^{16}$\\
 	8 & c-C$^{13}$CH$_{4}$O & 13 (5) & 34.9 - 64.0 & $5.32_{-0.58}^{+0.38}\times10^{15}$ & \multicolumn{5}{c}{2-Propenal}\\ 
 	     \multicolumn{5}{c}{Acetaldehyde} & 19 & C$_{2}$H$_{3}$CHO & 96 (44) & 46.7 - 410.4 & $7.71_{-0.72}^{+0.83}\times10^{15}$\\ 
    9 & CH$_{3}$CHO & 313 (74) & 18.3 - 775.9 & $6.70_{-1.44}^{+0.43}\times10^{17}$ & \multicolumn{5}{c}{Methyl cyanide}\\
 	  & CH$_{3}$CHO v$_{15}$=1 & 193 (54) & 224.6 - 1007.9 & $5.73_{-0.49}^{+1.05}\times10^{17}$ & 20 & CH$_{3}$CN & 19 (4) & 68.9 - 442.4 & $9.23_{-1.35}^{+0.51}\times10^{15}$\\  
      & CH$_{3}$CHO v$_{15}$=2 & 87 (47) & 439.5 - 975.3 & $3.97_{-0.46}^{+0.88}\times10^{17}$ &   & CH$_{3}$CN v$_{8}$=1 & 20 (7) & 588.0 - 679.2 & $2.32_{-0.42}^{+0.10}\times10^{16}$\\ 
    10 & $^{13}$CH$_{3}$CHO & 95 (67) & 70.2 - 828.0 & $1.82_{-0.25}^{+0.15}\times10^{16}$ & 21 & $^{13}$CH$_{3}$CN & 12 (9) & 78.0 - 167.3 & $5.23_{-0.27}^{+0.16}\times10^{14}$\\ 
 	   & $^{13}$CH$_{3}$CHO v$_{15}$=1 & 18 (11) & 286.1 - 366.9 & $1.84_{-0.46}^{+0.14}\times10^{16}$ & 22 & CH$_{3}$$^{13}$CN & 8 (4) & 80.3 - 156.9 & $4.86_{-0.15}^{+0.37}\times10^{14}$\\ 
    11 & CH$_{3}$$^{13}$CHO & 89 (54) & 71.1 - 275.2 & $1.74_{-0.14}^{+0.51}\times10^{16}$ & 23 & CH$_{2}$DCN & 24 (7) & 75.8 - 294.0 & $1.24_{-0.10}^{+0.04}\times10^{15}$\\ 
    \hline
    \enddata
    \tablenotetext{a}{The number of all identified lines over 3$\sigma$ noise level. In parentheses is the number of isolated, optically thin molecular lines used in the LTE line fitting.}
\end{deluxetable*}
    
To find clues about the formation pathways of the molecules, we organized the observed COMs in V883 Ori to see if their isomers were also found (see Table \ref{tb:isomers}). Among molecules listed in Table \ref{tb:isomers}, five isomers were undetected in V883 Ori. Propadiene, vinyl alcohol, and propylene oxide are detected only in limited astrophysical objects. Propadiene has only been revealed in the atmosphere of Titan \citep{Lombardo_2019}, and vinyl alcohol detection has been reported in the high-mass star-forming region Sagittarius B2(N) \citep{Turner_2001}, the starless core TMC-1 \citep{Agundez}, and the giant molecular cloud G+0.693-0.027 \citep{Jimenez-Serra}. Propylene oxide, an isomer of propanal and acetone, was discovered only in an extended molecular shell around the protostellar clusters in the star-forming region of Sagittarius B2 \citep{McGuire2016}.
On the other hand, the two isomers of methyl formate (glycolaldehyde and acetic acid) are more commonly detected than the above-mentioned COMs in various star-forming regions \citep{Hollis_2001, PILS2, Ceccarelli_PPVII}. However, despite the possibility of several emission lines of the two isomers within our observed frequency range, they were not detected.\par

\begin{deluxetable}{cccc}
    \centering
    \tabletypesize{\scriptsize}
    \tablecaption{Isomers of detected COMs in V883 Ori, with the detection information in other star-forming regions (SFRs).} \label{tb:isomers}
    \tablehead{\colhead{Formula} & \colhead{Isomer} & \multicolumn{2}{c}{Detection}\\
    & & \colhead{V883 Ori} & \colhead{SFRs}}
    \startdata
    \multirow{2}{*}{\centering C$_3$H$_4$} & CH$_3$CCH (Propyne) & Yes & Yes\\
                                & CH$_2$CCH$_2$ (Propadiene) & No & No\\\hline
    \multirow{3}{*}{\centering C$_2$H$_4$O} & c-C$_2$H$_4$O (Ethylene oxide) & Yes & Yes\\
                                            & CH$_3$CHO (Acetaldehyde) & Yes & Yes\\    
                                            & CH$_2$CHOH (Vinyl alcohol) & No & Yes\\\hline
    \multirow{2}{*}{\centering C$_2$H$_6$O} & CH$_3$OCH$_3$ (Dimethyl ether) & Yes & Yes\\
                                            & C$_2$H$_5$OH (Ethanol) & Yes & Yes\\\hline
    \multirow{3}{*}{\centering C$_3$H$_6$O} & s-C$_2$H$_5$CHO (Propanal) & Yes & Yes\\
                                            & CH$_3$COCH$_3$ (Acetone) & Yes & Yes\\
                                            & CH$_3$CHCH$_2$O (Propylene oxide) & No & Yes\\\hline
    \multirow{3}{*}{\centering C$_2$H$_4$O$_2$} & CH$_{3}$OCHO (Methyl formate) & Yes & Yes\\
                                     & CH$_{2}$OHCHO (Glycolaldehyde) & No & Yes\\
                                     & CH$_{3}$COOH (Acetic acid) & No & Yes\\\hline
    \enddata
\end{deluxetable}
    
\subsection{Column densities of the detected molecules} \label{sec:column density result}
\begin{figure*}[t]
    \centering
    \includegraphics[trim=0.2cm 0cm 0.2cm 0cm, clip=true,width=1\textwidth]{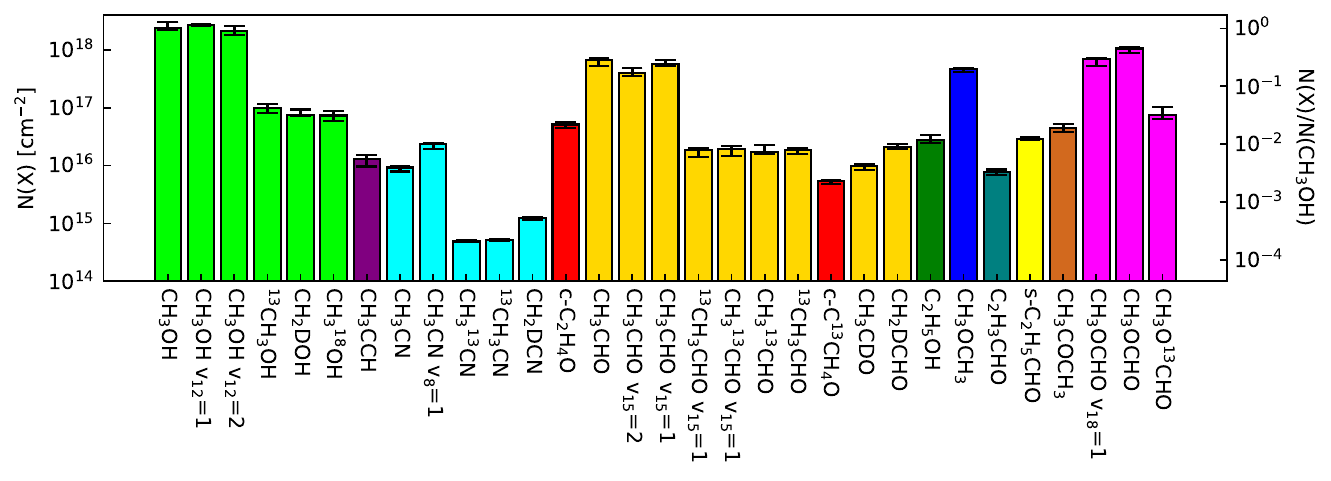}
    \caption{Bar plot for the (relative) column densities of all the COMs detected in V883 Ori. The column density values are from Table \ref{tb:identified}. COMs are sorted by increasing the value of mass, and the same color indicates isotopologues of the same species. The error bars are the uncertainties in the column densities, not the relative column densities.}
    \label{fig:coms barplot}
\end{figure*}

Figure \ref{fig:coms barplot} shows a bar plot displaying the column densities listed in Table \ref{tb:identified}. The column densities span several orders of magnitude, with values that fall into two main groups based primarily on main isotopologues: CH$_{3}$OH, CH$_{3}$CHO, CH$_{3}$OCH$_{3}$, and CH$_{3}$OCHO are the abundant species with column densities of 10$^{17}$$-$10$^{18}$ cm$^{-2}$, while CH$_{3}$CCH, CH$_{3}$CN, c-C$_{2}$H$_{4}$O, C$_{2}$H$_{5}$OH, C$_{2}$H$_{3}$CHO, s-C$_{2}$H$_{5}$CHO, and CH$_{3}$COCH$_{3}$ exhibit relatively small column densities of about 10$^{16}$ cm$^{-2}$.\par

The column densities obtained from transitions in a vibrationally excited state (e.g., CH$_{3}$OH v$_{12}$=1) represent the total number of the molecule (e.g., CH$_{3}$OH as a whole), not the number of molecules in that particular excited state. Therefore, the column densities estimated from the rotational transitions, both in the vibrationally ground state and in the excited state, must be consistent within fitting errors, like in the case of CH$_{3}$OH.\par

However, there are exceptions (CH$_{3}$OCHO v$_{18}$=1, CH$_{3}$CHO v$_{15}$=2, and CH$_{3}$CN v$_{8}$=1) that show differences in derived column densities by up to a factor of 2 when compared with their lower vibrational energy levels. The excitation to the higher vibrational energy states requires hotter and denser conditions. Therefore, the transitions in vibrationally excited states may originate from hotter regions of the disk such as its upper atmosphere. In Appendix \ref{sec:temperature variation}, we test hotter temperature conditions to show that the column densities of the vibrationally excited species derived from a higher temperature of 200 K better match with the 120 K derived column densities of the ground state species. In addition, we demonstrate that a fixed rotation temperature of 120 K is more appropriate than higher temperatures (200 K, 300 K) for most species.\par
    
\section{Discussion} \label{sec:discussion}
\subsection{Comparions with other sources}
This section presents comparisons of the chemical composition of several different young stellar objects, as well as a Solar System comet. A comparison of the abundances relative to methanol (i.e., N(X)/N(CH$_3$OH)) of all main isotopologues detected in V883 Ori, along with the upper limits for some undetected species, against two representative sources is presented in Section \ref{sec:comparison against hot corino and comet}. A comparison with a number of sources, but for a few molecules, is presented in Section \ref{sec:comparison against various sources}. The comparison results show that V883 Ori disk exhibits unique characteristics that set it apart from other targets.

\subsubsection{Comparisons against a hot corino and a comet}\label{sec:comparison against hot corino and comet}
We compare the COMs composition of the V883 Ori disk against the compositions of a Class 0 hot corino IRAS 16293-2422 and a Solar System comet 67P/Churyumov–Gerasimenko (hereafter 67P/C-G), each representing, respectively, an earlier and more evolved stage than V883 Ori.\par

IRAS 16293-2422 is a Class 0 low-mass protostellar binary system (IRAS 16293-2422A and IRAS 16293-2422B) with a projected separation of 747 au. It is located in the Ophiuchus cloud complex at a distance of 141 pc \citep{PILS2, Dzib2018, PILS19}. IRAS 16293-2422 has been studied with an unbiased ALMA spectral survey, Protostellar Interferometric Line Survey (PILS, \citet{PILS2}), which covers about 35 GHz wide frequency range in Band 7. Various COMs were detected by identifying over 10,000 lines in this Class 0 hot corino \citep{jorgensen2020_araa}. Thus, the PILS survey is a good reference for comparing chemical properties in detail with this study.\par

In Appendix \ref{sec:long table} we tabulate a detailed comparison of the column densities of COMs and their column density relative to CH$_3$OH between V883 Ori and IRAS 16293-2422 (Table \ref{tb:comparisons}). For this comparison, we used the results of IRAS 16293-2422B in the IRAS 16293-2422 binary system. For 40 species including isotopologues that are not detected in V883 Ori, we present the upper limits of column densities, which are those column densities when their maximum intensities of the model spectra correspond to the 3$\sigma$ levels of the observations. \par

Overall, more diverse main isotopologues of COMs were detected in IRAS 16293-2422B than V883 Ori. In addition, all the main COMs detected in V883 Ori are also present in IRAS 16293-2422B. Table \ref{tb:not detected coms} lists COMs found only in IRAS 16293-2422B with non-detection in V883 Ori. Glycolaldehyde and its isomer, acetic acid, and its reduced alcohol, ethylene glycol, were not detected in V883 Ori. Most of them are molecules containing the hydroxymethyl group (CH$_{2}$OH). In addition, the nitrogen-bearing molecules, except for CH$_3$CN, are also missing in V883 Ori. We discuss implications of the non-detections of CH$_{2}$OH-, nitrogen-bearing COMs in Section \ref{sec:implication}.\par

\begin{figure*}[t]
    \centering
    \includegraphics[trim=0.1cm 0cm 0.1cm 0cm, clip=true,width=0.495\textwidth]{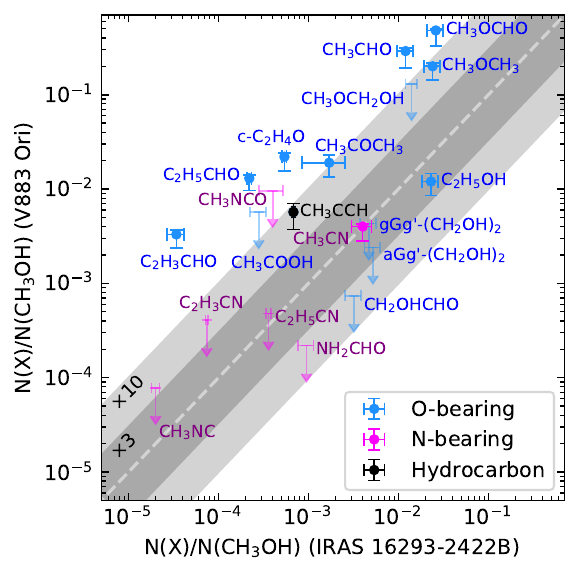}
    \includegraphics[trim=0.1cm 0cm 0.1cm 0cm, clip=true,width=0.495\textwidth]{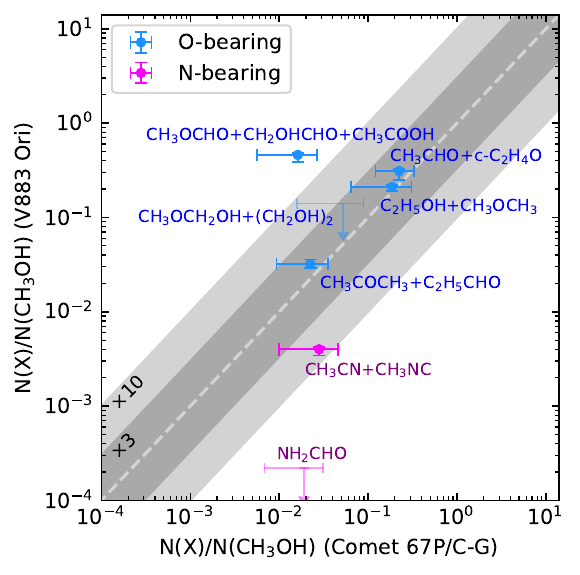}
    \caption{Comparison of the column density of COMs relative to methanol between V883 Ori $\&$ IRAS 16293-2422B (left panel, based on Table \ref{tb:comparisons}), and V883 Ori $\&$ Comet 67P/Churyumov-Gerasimenko (right panel). The comet data represent relative bulk compositions of species with the same mass, and are taken from \citet{Schuhmann2019, Rubin2019}. For the non-detected molecules in V883 Ori, the upper limits are induced from the 3$\sigma$ levels. The line representing 1:1 is indicated by a gray dashed line. The lines representing 3 times and 10 times are shown as gray solid lines with dark and light gray shaded areas, respectively.}
    \label{fig:comparison_PILS_comet}
\end{figure*}

We plot a diagram for abundances relative to methanol to compare these two sources (Left panel of Figure \ref{fig:comparison_PILS_comet}), along with upper limits of the species listed in Table \ref{tb:not detected coms}. The oxygen-bearing COMs in V883 Ori are generally more abundant than those in IRAS 16293-2422B, with abundances more than an order of magnitude higher. The molecule with the largest difference is 2-propenal (C$_{2}$H$_{3}$CHO), which shows two orders of magnitude difference. However, ethanol (C$_{2}$H$_{5}$OH) shows a smaller abundance in V883 Ori than IRAS 16293-2422B, by about a factor of two. Among the undetected molecules in V883 Ori, the upper limit of glycolaldehyde (CH$_{2}$OHCHO) and formamide (NH$_{2}$CHO) are several times less than the abundances of IRAS 16293-2422B.
The only detected nitrogen-bearing molecule in V883 Ori is methyl cyanide (CH$_{3}$CN), whose abundance is comparable with IRAS 16293-2422B. Therefore, the abundances of the detected COMs in the V883 Ori disk are overall higher than those of the Class 0 hot corino IRAS 16293-2422B, still with a positive correlation. However, ethanol and methyl cyanide deviate from the correlation and have similar abundances in these two targets.\par
    
We also make a comparison of the compositions with Comet 67P/C-G, a Jupiter family comet visited by the Rosetta Mission. Through the mission, various organic compounds were discovered, and their abundances were determined \citep{Schuhmann2019, Rubin2019}.
The cometary values were presented and compared with the abundances of Class 0 hot corino IRAS 16293-2422B in a study by \citet{PILS19}. They revealed that the CHO-, nitrogen-bearing COMs in 67P/C-G have higher abundances systematically, but still with a good correlation between the two sources.\par

The right panel of Figure \ref{fig:comparison_PILS_comet} shows the abundances of COMs detected in V883 Ori and 67P/C-G. Abundances in the two objects are consistent without any obvious offset, although the CH$_{3}$OCHO and CH$_{3}$CN abundances in V883 Ori are higher and lower, respectively, by an order of magnitude than in the comet.\par

The higher abundances than the Class 0 hot corino and similar abundances with the cometary ices may indicate the possibility of in-situ chemical evolution in a disk around a young star rather than maintaining the original compositions from the protostellar envelope.

\begin{deluxetable}{cccc}
    \tabletypesize{\scriptsize}
    \tablecaption{COMs undetected in V883 Ori but detected in IRAS 16293-2422B. 
    \label{tb:not detected coms}
    }
    \tablehead{\colhead{Species} & \colhead{Formula} & \colhead{Species} & \colhead{Formula}}
    \startdata
    Acetic acid & CH$_{3}$COOH & Methyl isocyanide & CH$_{3}$NC\\
    Glycolaldehyde & CH$_{2}$OHCHO & Formamide & NH$_{2}$CHO\\
    Ethylene glycol & (CH$_{2}$OH)$_2$ & Methyl isocyanate & CH$_{3}$NCO\\
    Methoxymethanol & CH$_{3}$OCH$_{2}$OH & Vinyl cyanide & C$_{2}$H$_{3}$CN\\
     &  & Ethyl cyanide & C$_{2}$H$_{5}$CN\\
    \hline
    \enddata
\end{deluxetable}

\subsubsection{Comparisons against various sources}\label{sec:comparison against various sources}
For a few COMs, we show an additional comparison with various studies of Class 0/I hot corinos, including the ALMA/PILS \citep{PILS2}, IRAM/CALYPSO \citep{Belloche2020}, ALMA/PEACHES \citep{Yang_2021}, ALMA/ALMASOP \citep{Hsu_2022}, and the ALMA observation of HOPS 373SW, which is a very young hot corino \citep{Lee_HOPS373}, and studies for disks such as HH 212 (Class 0 disk atmosphere, \citealt{Lee_Chin_Fei2019}), Oph IRS 48 (transition disk, \citealt{Brunken2022}), and V883 Ori (Class I/II disk including disk midplane, \citealt{jelee19, yamato2024chemistry}). We also compare with high-mass protostars: Sgr B2(N2) by ALMA/EMoCA \citep{Belloche, muller2016, Bonfand2019, Ordu2019} and various hot cores by ALMA/CoCCoA \citep{coccoa2023} and by \citet{Giseon2022}.\par

\begin{figure*}
\includegraphics[trim=0cm 0cm 0cm 0cm, clip=true,width=1\textwidth]{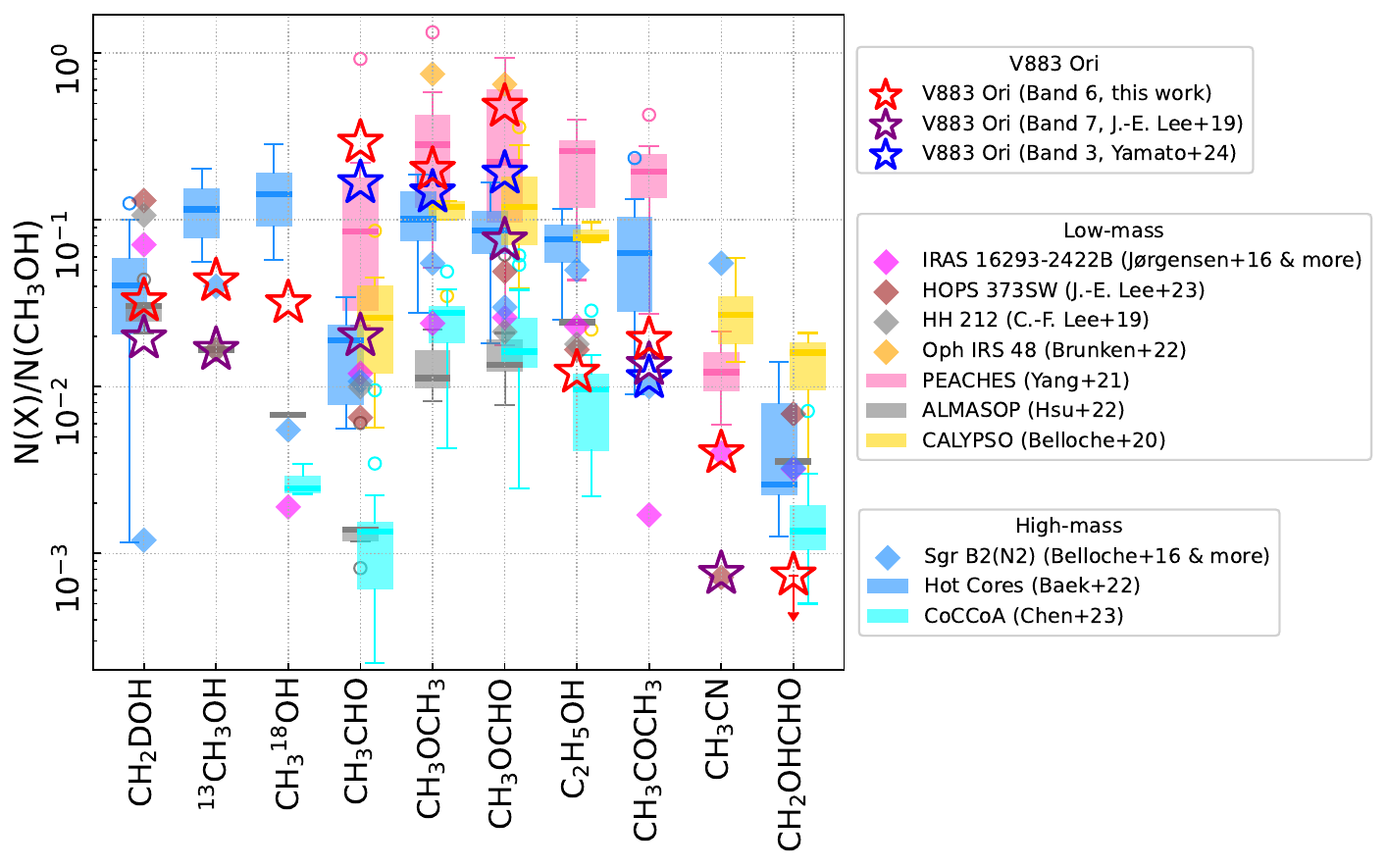}
\caption{Column densities of COMs relative to methanol for various hot corinos/cores studies. There are three types of markers: stars for three V883 Ori studies, diamonds for five studies of a single source, and boxplots for five survey studies. The boxes of the boxplots extend across the first (Q1) to the third (Q3) quartile of the data and have a line at the median. The whiskers extend from the box to the farthest data point that is within 1.5 times the Q3$-$Q1 range from the box. The outliers of the distribution are shown as open circles. In legends, we separate the V883 Ori, hot corinos, and hot cores studies into three groups. For CH$_{2}$OHCHO the upper limit of V883 Ori is marked.}
\label{fig:abund_comp}
\end{figure*}

Figure \ref{fig:abund_comp} shows the abundances of COMs relative to methanol derived from the aforementioned studies. For each study, the column density of methanol that emits optically thick lines was obtained in the following manners. CH$_{3}$$^{18}$OH was used for PILS and CoCCoA. CALYPSO and PEACHES surveys, the study of IRS 48, the hot cores \citep{Giseon2022}, and V883 Ori (from \citealt{yamato2024chemistry} and this work) directly derived the column densities of methanol using its main isotopologue. ALMASOP survey, the study of V883 Ori (from \citealt{jelee19}), HOPS 373SW and HH 212 took the values calculated from 60$\times$$^{13}$CH$_{3}$OH. In fact, the study of ALMASOP and HH 212 used 50$\times$$^{13}$CH$_{3}$OH, but we multiplied their values by 60/50 to make the comparison consistent.\par

First, we compare the results of the V883 Ori studies that used different ALMA Bands: Band 7 from \citet{jelee19}, Band 3 from \citet{yamato2024chemistry}, and Band 6 from this work (stars in Figure \ref{fig:abund_comp}), which are all corrected for the dust optical depth. The column densities of methanol in the three studies are similar as  3.2$\times$10$^{18}$ cm$^{-2}$ ($=60\times$$^{13}$CH$_{3}$OH), 3.9$\times$10$^{18}$ cm$^{-2}$, 2.3$\times$10$^{18}$ cm$^{-2}$, respectively. On the other hand, the relative column densities of CH$_{3}$CHO, CH$_{3}$OCHO, and CH$_{3}$CN derived from the Band 7 study are underestimated by about 5 to 14 times, while the Band 3 study and this work are consistent within about a factor of two. This can be attributed to the line optical depth effect, which is examined in Section \ref{sec:optical depth effect}, because both the Band 3 study and this work used only optically thin lines to estimate column densities. 
 
Therefore, it is important to take into account the line optical depth effect when estimating column densities. Otherwise, there is the potential to be underestimated several times.\par

Figure \ref{fig:abund_comp} shows a large spread in abundances across about four orders of magnitude without significant systematic differences between low-mass and high-mass protostars, as previously revealed (e.g., \citet{Giseon2022, coccoa2023}). Although V883 Ori falls within these ranges, it shows some distinctive trends that make it different from other sources.
Abundances of CH$_{3}$CHO, CH$_{3}$OCH$_{3}$, and CH$_{3}$OCHO in V883 Ori tend to be higher than most of the sources. The other disk source, Oph IRS 48 \citep{Brunken2022}, shows higher values than V883 Ori, although only two molecules were compared. Also, some of PEACHES samples, including outliers, show higher abundances, which may be caused by their underestimated CH$_{3}$OH column densities, as mentioned by \citet{Yang_2021}.\par

On the other hand, C$_{2}$H$_{5}$OH, CH$_{3}$COCH$_{3}$, and CH$_{3}$CN in V883 Ori are less abundant (or comparable) than (to) those of other objects. Also, the upper limit of CH$_{2}$OHCHO abundance is almost the lowest.
The column density ratios using the upper limits of CH$_{2}$OH-bearing COMs verify the lack of them clearly. For example, a lower limit of the ratio between two isomers of C$_{2}$H$_{4}$O$_{2}$, CH$_{3}$OCHO/CH$_{2}$OHCHO is 647 for V883 Ori, while the average ratio of the other sources in Figure \ref{fig:abund_comp} is 15.9. Similarly, a lower limit of CH$_{3}$OCHO/(CH$_{2}$OH)$_{2}$ is 72 (see Table \ref{tb:comparisons}). Both values are the highest when compared to the column density ratios in various star-forming regions listed in \citet{Lykke2015} and \citet{Mininni2020}. Furthermore, the average column density ratio of CH$_{3}$CHO/CH$_{2}$OHCHO of the sources in Figure \ref{fig:abund_comp} is 2.4, while the lower limit of this ratio in V883 Ori is 394.

HOPS 373SW \citep{Lee_HOPS373}, HH 212 \citep{Lee_Chin_Fei2019}, and the sources in the ALMASOP survey \citep{Hsu_2022} are Class 0/I protostellar sources located in the Orion clouds. Thus, they could have similar initial chemical compositions. However, V883 Ori exhibits distinctive chemical features compared to these other Orion sources. In addition, the composition of COMs detected in the disk atmosphere of HH 212 is also different from the V883 Ori disk. Therefore, this may suggest that the chemistry evolved in the deeper layers of the disk, or the physical and/or dynamic environment in V883 Ori was different.\par

Lastly, we note that the column densities derived in this study with a fixed beam-filling factor are likely lower limits (Appendix \ref{sec:effect of parameters}). We use the beam-filling factor derived from the continuum fitting, which is consistent with the value derived from the saturated CH$_{3}$OH lines. However, the emitting area of other species might be smaller than CH$_{3}$OH. Nonetheless, this would not affect our conclusions (e.g., the extremely high lower limit of CH$_{3}$OCHO/CH$_{2}$OHCHO ratio in V883 Ori).

\subsection{Distinct chemical characteristics of the V883 Ori disk} \label{sec:implication}
As mentioned above, V883 Ori is notable for lack of CH$_{2}$OH- and nitrogen-bearing COMs, although the other oxygen-bearing COMs such as CH$_{3}$CHO, CH$_{3}$OCH$_{3}$, and CH$_{3}$OCHO are rather abundant compared to various other sources. Likewise, in this section, we discuss the implications of these phenomena.\par 

We focus on the grain-surface reactions rather than the gas-phase chemistry because the detected COMs had been trapped in the ice mantles in the quiescent phase and were sublimated by the recent outburst event. Also, the time scale of gas-phase reactions is an order of 10$^{4}$ years, which is longer than the duration of the outburst and the freeze-out time scale \citep{Nomura2009, jelee19}. While \citet{Taquet2016} showed that gas-phase formation of COMs could be important even in an outburst of 100 yr timescale, the key molecule in their model, ammonia (NH$_{3}$), has not been detected in V883 ori (Tobin et al. in IAU symposium 2023). Hence, we reemphasize the fact that V883 Ori is an excellent target for studying the ice composition of a disk, which sets the initial conditions of planet formation.\par

\subsubsection{COMs formation on water-rich ice environment}
One of the important conditions in COM chemistry is the main ice component covering grain surfaces where chemical reactions occur. Although COMs have been believed to be initiated by the hydrogenation of CO on grain surfaces \citep{Watanabe_2002, Garrod_2013, Chuang_2018} in the cold ($<10$ K) prestellar cores, they could also form in warmer water-rich ice mantles \citep{Bergner_2017, Lamberts_2017, Qasim_2018, Potapov_2021}, where OH-radicals can be produced by photolysis of water (H$_{2}$O + h$\nu$ $\longrightarrow$ OH + H).\par

In environments where the temperature exceeds 10 K, desorption of CO and H atoms becomes more efficient. This reduces the likelihood of high CO coverage on the ice surface, inhibiting its hydrogenation reactions. As a result, surface chemistry may become dominated by OH radicals, which are produced by water photodissociation. For example, methanol (CH$_3$OH) can form via reactions such as ${\rm CH_4 + OH \longrightarrow CH_3 + H_2O}$ and ${\rm CH_3 + OH \longrightarrow CH_3OH}$ \citep{Qasim_2018}. Ice surface reactions driven by OH, rather than hydrogenation, could play a key role in explaining the deficiency of CH$_2$OH-bearing COMs in V883 Ori.\par

The OH-radicals can react with neighbouring methanol on the water ice surface to form CH$_{3}$O-radicals from the following reaction \citep{Shannon2013, Ishibashi_2021}:
\begin{equation}
    \label{eq:OH}
    {\rm CH_3OH+OH \longrightarrow CH_3O+H_2O.}
\end{equation}
On the other hand, CH$_{2}$OH, which is the structurally isomeric radical of CH$_{3}$O, can be made by photodissociation of methanol \citep{Garrod_2008}, and by isomerization of CH$_{3}$O \citep{TACHIKAWA199327}:
\begin{equation}
    {\rm CH_3OH+h\nu \longrightarrow CH_2OH\ or\ CH_3O+H,}
\end{equation}
\begin{equation}
    {\rm CH_3O+CH_3OH \longrightarrow CH_3OH+CH_2OH.}
\end{equation}
\par

However, in the OH-rich environment where water is more abundant than methanol, Reaction \ref{eq:OH} occurs predominantly \citep{Ishibashi_2021, Ishibashi2024}. Furthermore, the two isomers interact with the water ice surface differently, resulting in different reactivity. For example, CH$_{2}$OH reacts with hydrogen more easily than CH$_{3}$O and is therefore consumed more efficiently \citep{Enrique2022}. As a result, the abundance ratio of CH$_{3}$O/CH$_{2}$OH in water ice would be higher compared to CO-, and thus, methanol-dominated ice \citep{Ishibashi_2021, Enrique2022, Ishibashi2024}.\par

This abundance difference between the two isomeric radicals could presumably yield large amounts of CH$_{3}$O-derived products, such as methyl formate (CH$_{3}$OCHO) and dimethyl ether (CH$_{3}$OCH$_{3}$), while producing relatively small amounts of CH$_{2}$OH-derived products, such as glycolaldehyde (CH$_{2}$OHCHO) and ethylene glycol ((CH$_{2}$OH)$_{2}$).\par

Furthermore, laboratory studies \citet{Chuang2020} showed an important surface chemistry for acetaldehyde on the water-rich C$_{2}$H$_{2}$ ice, where CO molecules are not frozen on to the surface of dust grains, as described below.\par

The widely available OH-radicals can actively react with the initial C$_{2}$H$_{2}$ through reaction
\begin{equation}
    {\rm C_2H_2+OH \longrightarrow CHCHOH.}
\end{equation}
Then the CHCHOH-radical combines with hydrogen, forming vinyl alcohol (CH$_{2}$CHOH),
\begin{equation}
    {\rm CHCHOH+H \longrightarrow CH_2CHOH.}
\end{equation}
These vinyl alcohol molecules eventually isomerize to acetaldehyde (CH$_{2}$CHOH $\longleftrightarrow$ CH$_{3}$CHO) over the star-formation time scale because acetaldehyde has a lower potential energy than vinyl alcohol \citep{Chuang2020}.\par

The water-rich environment can be developed when the temperature is high enough to prevent CO from being frozen on grain surfaces \citep{Dishoeck2014}. Moreover, \citet{Kouchi2021} reported that CO can be crystallized on icy grains, with the result that the water ice mantle is not fully covered by CO. Then, CH$_{3}$OH can easily react with water molecules. The time scale of crystallization of CO on water ice becomes shorter as the grain temperature rises \citep{Kouchi2021}.\par

Based on the D/H ratio of COMs detected in ALMA Band 3, \citet{yamato2024chemistry} suggested that COMs form on the lukewarm ($\sim$30-50 K) grain surfaces within its disk. The D/H ratio of molecules is an important indicator of the formation temperature of those molecules \citep{Lee_2015}.
In addition, \citet{Hoff2018, Hoff2020} suggested that the young protostellar disks tend to be warmer than the more evolved Class II disks, which may prevent CO freeze-out onto the dust surface. Therefore, these previous studies of COMs in disks consistently support the hypothesis that the COMs in the disk around V883 Ori formed in a water-rich, lukewarm grain environment.\par 

An interesting phenomenon may provide insight into the distinct chemical environment of V883 Ori; the H$_{2}$O/CH$_{3}$OH ratio ($\sim$13) is about an order of magnitude lower compared to IRAS 16293-2422 (330) and Comet 67P (476) \citep{PILS19}. The water column density was calculated from our estimated HDO column density (6.68$\times$10$^{16}$ cm$^{-2}$) and the disk-averaged HDO:H$_{2}$O ratio (2.26$\times$10$^{-3}$) reported by \citet{Tobin2023}. The high methanol abundance relative to water may provide constraints on the chemical processes occurring in the water-rich ice environment inferred for V883 Ori. Further exploration with chemical models will be essential for a more comprehensive understanding.

\subsubsection{Nitrogen-bearing COMs}
Another intriguing characteristic of COMs in V883 Ori is that nitrogen-bearing COMs, except for methyl cyanide (CH$_{3}$CN), are missing (\citet{jelee19, yamato2024chemistry}, and this work). The only detected nitrogen-bearing COM, CH$_{3}$CN, also has a low abundance compared to other sources (see Figure \ref{fig:abund_comp}). For the oxygen-bearing COMs, only CH$_{2}$OH-bearing species are exclusively absent, while the nitrogen-bearing COMs are missing regardless of their containing radicals. For example, NH$_{2}$CHO (NH$_{2}$-radical), CH$_{3}$NCO (NCO-radical), C$_{2}$H$_{3}$CN (CN-radical), and C$_{2}$H$_{5}$CN (CN-radical), which were detected in IRAS 16293-2422B, were not detected in V883 Ori (Table \ref{tb:not detected coms}).\par

This distinct chemical phenomenon suggests that nitrogen-bearing COMs detected in hot cores and hot corinos \citep{Mininni2023, Taniguchi_2023} may have formed via gas-phase chemistry following the sublimation of ammonia (NH$_{3}$) from ice mantles into the gas phase. Given that the time scale for the gas-phase chemistry of COMs \citep{Nomura2009} is much longer than the time scale of the outburst event, the duration following the burst in V883 Ori may not be sufficient for the formation of nitrogen-bearing COMs.\par

Before the burst, ammonia, present in a water-rich ice mixture (e.g., H$_{2}$O:CO:NH$_{3}$), is expected to have easily formed NH$_{2}$CHO \citep{Chuang2022} and sublimated during the burst. However, in V883 Ori, NH$_2$CHO showed only a low upper limit of abundance, suggesting that ammonia did not participate in the formation of NH$_2$CHO. This nondetection of NH$_{2}$CHO may imply that most nitrogen is sequestered in refractory components, such as ammonium salts within the disk, and that nitrogen-bearing COMs rarely form in this environment \citep{Boogert2015, oberg2023, yamato2024chemistry}.\par

This hypothesis aligns with observations of comets, which are partly composed of ice from the disk midplane and also show a depletion of nitrogen-bearing molecules. \citet{Altwegg2020a} suggested that nitrogen in Comet 67P/C–G is primarily trapped as ammonium salts  (e.g., NH$_{4}$$^{+}$Cl$^{-}$, NH$_{4}$$^{+}$CN$^{-}$, NH$_{4}$$^{+}$OCN$^{-}$, NH$_{4}$$^{+}$HCOO$^{-}$, and NH$_{4}$$^{+}$CH$_{3}$COO$^{-}$). More recently, NH$_{4}$$^{+}$SH$^{-}$ was identified as the most abundant ammonium salt \citep{Altwegg2022}.

Ammonium salts have higher sublimation temperatures than COMs, meaning that their base (ammonia) or acids (e.g., HCl, HNCO, HCN, H$_2$S) may exist closer to the inner disk regions. However, the high dust opacity toward the central region complicates testing this hypothesis. To better explore the nitrogen reservoir in V883 Ori, observations at wavelengths longer than those used in ALMA Band 3 \citep{yamato2024chemistry} will be essential. Additionally, since V883 Ori is the only disk source where various COMs have been detected in the disk midplane, further observations of additional disk sources are needed to confirm whether disk midplanes are consistently barren of nitrogen-bearing COMs.

\subsection{Carbon isotope ratios of COMs}\label{sec:12C/13C}
Isotope fractionation occurs under specific conditions, such as cold regions and/or intense UV radiation (e.g., \citealt{Lyons2018, Nomura2023}). Isotope ratio between $^{12}$C- and $^{13}$C- bearing isotopologues can be fractionated through the isotope-selective photodissociation of CO \citep{Visser2009} or the isotope exchange reactions such as $\mathrm{^{13}C}^+ + \mathrm{CO} \rightleftharpoons \mathrm{C}^+ + \mathrm{^{13}CO} + 35 \mathrm{K}$ \citep{Langer1984, Woods_2009, Loison2020}. In this section, we derive the carbon isotope ratio of COMs based on the column density estimates in Section \ref{sec:column density result}.\par

\begin{figure}[t]
    \centering
    \includegraphics[trim=0.2cm 0cm 0.2cm 0cm, clip=true,width=0.48\textwidth]{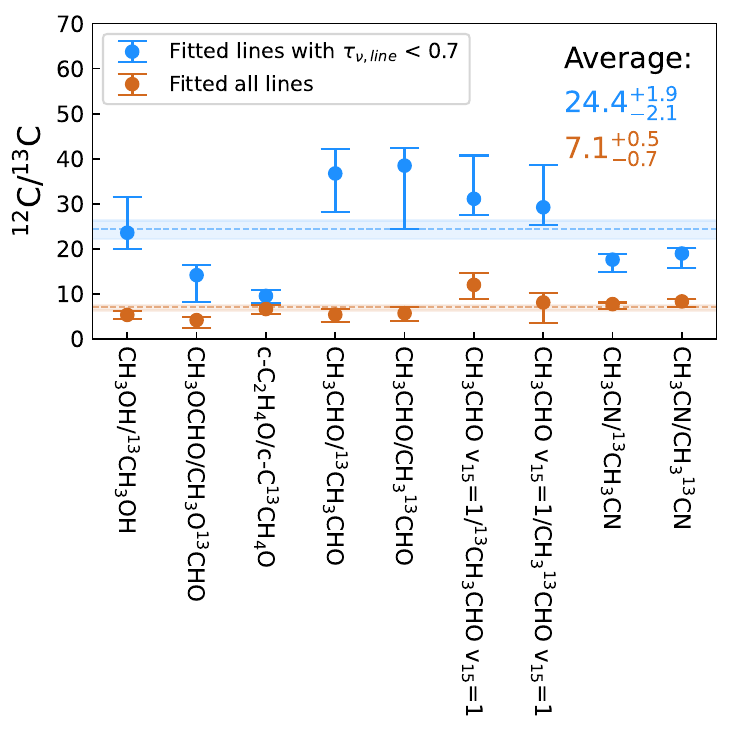}
    \caption{$^{12}$C/$^{13}$C isotopic ratios. The column density results using isolated lines of $\tau$$_{\rm \nu,line}$ smaller than 0.7 (Table \ref{tb:identified}) and using isolated lines of all $\tau$$_{\rm \nu,line}$ are plotted respectively. The dashed lines represent the average $^{12}$C/$^{13}$C values and the shaded areas show the 1$\sigma$ error range for the average values.}
    \label{fig:C_isotopic_ratio}
\end{figure}

Figure \ref{fig:C_isotopic_ratio} shows $^{12}$C/$^{13}$C ratios that are calculated from every species whose carbon isotopologues are detected in V883 Ori. The ratios are calculated by dividing pure column densities without any statistical correction. The blue circles indicate the ratios using only optically thin lines ($\tau_{\rm \nu,line} <$ 0.7), while the brown circles denote ratios calculated using all lines. The average isotopic ratio derived from only optically thin lines is 3.4 times higher than that derived from all lines because many lines of the $^{12}$C-bearing isotopologues are optically thick, and thus, their column densities could be underestimated, as seen in the Section \ref{sec:optical depth effect}.\par

As a result, in V883 Ori, the COMs have the $^{12}$C/$^{13}$C ratio of 24.4 on average (from minimum 9.6 to maximum 38.5), which is distinct from the local ISM values \citep[$\sim$50--70,][]{Kahane_2018,Langer1993,Milam_2005} and the value derived from methanol in the Comet 67P/C-G \citep[91,][]{Altwegg2020b}. However, this low ratio is consistent with the value $\sim20-30$ presented by \citet{yamato2024chemistry}. It is also similar to the results of the PILS survey, with 17 for dimethyl ether and 27 for glycolaldehyde \citep{PILS2, PILS14}. HOPS 373SW also shows a low carbon isotopic ratio if CH$_{3}$OH v$_{12}$=2 is employed for the calculation of a methanol abundance \citep{Lee_HOPS373}. Furthermore, $^{12}$CH$_3$OH/$^{13}$CH$_3$OH isotope ratios derived from several hot cores range from 5 to 20 \citep{Giseon2022}. Therefore, the carbon isotope fractionation of COMs occurs in general during the star formation process, although its origins have not been thoroughly investigated yet \citep{PILS2, PILS14}.

\section{Summary} \label{sec:summary}
Utilizing our unbiased spectral survey of a low-mass, eruptive young star V883 Ori in ALMA Band 6 over the frequency coverage of $\sim$55 GHz (220.7 to 274.9 GHz), we analyzed freshly sublimated Complex Organic Molecules (COMs) in the Keplerian disk around V883 Ori, which is believed to have erupted $\sim$130 years ago and has not had time to modify the initial ice chemical composition. Throughout the analysis, we used the disk-averaged spectra, which were extracted using the first principal component (PC1) filtering method. This is the first study that fully investigated freshly sublimated COMs in a disk of a young stellar object using unbiased line survey data. The main findings of this study are summarized as follows: 

\begin{enumerate}
    \item We examined every possible species in the spectral range ($\sim$1,500 species) through a systematic method and robustly discovered 23 COMs, including isotopologues, by identifying about 3,700 molecular emission lines ($\sim$80$\%$ of detected lines). These are mostly oxygen-bearing COMs. On the other hand, CH$_{2}$OH- and nitrogen-bearing COMs, except for methyl cyanide, are missing in V883 Ori.
    
    \item The disk-averaged spectra contain numerous optically thick molecular lines. Therefore, an iterative LTE line fitting method was used to properly derive the column densities of detected COMs, adopting only optically thin lines. Through this method, we showed that the column density of some molecules (e.g., methyl formate) could be underestimated by a factor of a few unless only optically thin lines were adopted for the line-fitting process. Thanks to our robust line identification over the abnormally wide frequency range, we could collect a sufficient number of optically thin, isolated lines for each species.

    \item Compared to various Class 0/I hot corinos, V883 Ori shows extremely high CH$_{3}$OCHO:CH$_{2}$OHCHO, CH$_{3}$CHO:CH$_{2}$OHCHO, and CH$_{3}$OCHO:(CH$_{2}$OH)$_{2}$ ratios. The total budget of carbon and oxygen within complex molecules is concentrated on the detected COMs rather than spreading out over more diverse species.

    \item Compared to another disk source (Oph IRS 48) and a Solar System comet (67P/C-G), the abundances of the detected COMs seem comparable.

    \item The COMs in the V883 Ori disk may have formed in a water-rich ice environment. The OH-radicals, which are widely available in the water-rich ice, can cause significant differences in abundance between CH$_{3}$O- and CH$_{2}$OH-radicals. They can also be used efficiently for OH-radical addition chemical reactions, resulting in the deficiency of CH$_{2}$OH-bearing COMs in V883 Ori.

    \item All nitrogen-bearing COMs are missing in V883 Ori with the exception of CH$_3$CN, which has a low abundance compared to other COM-rich sources. According to previous observations of the Solar System comet (67P/C-G), nitrogen might be trapped as ammonium salt in the dust grains.

    \item The average $^{12}$C/$^{13}$C ratio estimated from various COMs detected in V883 Ori is 24.4, which is comparable to the ratios calculated from the other hot corino/core sources but lower by $\sim2-4$ times when compared to the ratios in the Comet 67P/C-G and the local ISM.
\end{enumerate}

\section{Acknowledgements}
We greatly appreciate the thorough and constructive review by the anonymous referees. This work was supported by the New Faculty Startup Fund from Seoul National University and the NRF grant funded by the Korean government (MSIT) (grant numbers 2021R1A2C1011718 and RS-2024-00416859). JHJ was supported by the National Research Foundation of Korea (NRF) grant (NRF- 2019R1A2C2010885). GB was supported by Basic Science Research Program through the National Research Foundation of Korea (NRF) funded by the Ministry of Education (RS-2023-00247790). Y.A. acknowledges support by MEXT/JSPS Grants-in-Aid for Scientific Research (KAKENHI) Grant Numbers JP20H05847 and JP24K00674, and NAOJ ALMA Scientific Research Grant code 2019-13B. DJ is supported by NRC Canada and by an NSERC Discovery Grant. LC acknowledges financial support from ANID FONDECYT grant \#1211656.\par

This paper makes use of the following ALMA data: ADS/JAO.ALMA\#2019.1.00377.S. ALMA is a partnership of ESO (representing its member states), NSF (USA) and NINS (Japan), together with NRC (Canada), MOST and ASIAA (Taiwan), and KASI (Republic of Korea), in cooperation with the Republic of Chile. The Joint ALMA Observatory is operated by ESO, AUI/NRAO and NAOJ.\par
\begin{itemize}
    \item \it{Facility}: \rm ALMA \par
    \item \it{Software}: \rm Numpy \citep{Harris2020}, Scipy \citep{Virtanen2020}, Astropy \citep{astropy2013, Price-Whelan_2018, astropy2022}, Pandas \citep{scipy2020}, Matplotlib \citep{matplotlib}, XCLASS \citep{Moller17}
\end{itemize}

\bibliographystyle{aasjournal}
\bibliography{main}{}

\appendix
\section{Rotation diagram analysis of COMs}\label{sec:rotation diagram}
To derive the column densities of COMs, we applied iterative LTE line fitting in the main paper. For comparisons, we also adopted rotation diagrams, the most commonly used LTE analysis of molecular excitation, and here, we present the results. We used the same spectra as used for line fitting analysis, i.e., the PC1-filtered spectra corrected for dust attenuation, adopting the same beam-filling factor of 0.384. 

\subsection{Method}\label{sec:rotation diagram method}
Assuming all lines are optically thin, the column density and the rotation temperature of a molecular species can be inferred using the equation below \citep{Goldsmith_1999};
\begin{equation}\label{eq:rot}
        \ln\left(\frac{N_{\rm i}}{g_{\rm i}}\right)=\ln\left(\frac{N}{Q}\right)-\frac{E_{\rm i}}{kT_{\rm rot}}.
\end{equation}
The following equation can calculate the upper-level column density ($N_{\rm i}$) relevant to each line;

\begin{equation}\label{eq:rot2}
        N_{\rm i}=\frac{8\pi k \nu^{2}}{h c^{3} A_{\rm ij}} \int T_{\rm a} \,dv.
\end{equation}

Here, the $g_{\rm i}$ is the degeneracy and $E_{\rm i}$ is the energy of the upper level. Q is the partition function, which depends on the rotation temperature ($T_{\rm rot}$). $A_{\rm ij}$ is the Einstein A coefficient and $\nu$ is the frequency of the emission line. This spectroscopic information can be obtained from CDMS and JPL molecular databases queried through Splatalogue\footnote{https://splatalogue.online/}.\par

From the line identification result (Section \ref{sec:line identification result}), we selected isolated lines using the criterion of three times the channel width. Each line was fitted with a Gaussian profile to get the integrated intensity. For each molecule, the lines with large line widths, probably due to blending with other lines that are currently unidentified, are removed from the rotation diagram analysis because they tend to have bad fitting results. Only the lines of the low 80\% in the linewidth distribution were used.\par

\begin{figure*}[t]
    \centering
    \includegraphics[trim=0.1cm 0cm 0cm 0cm, clip=true,width=0.325\textwidth]{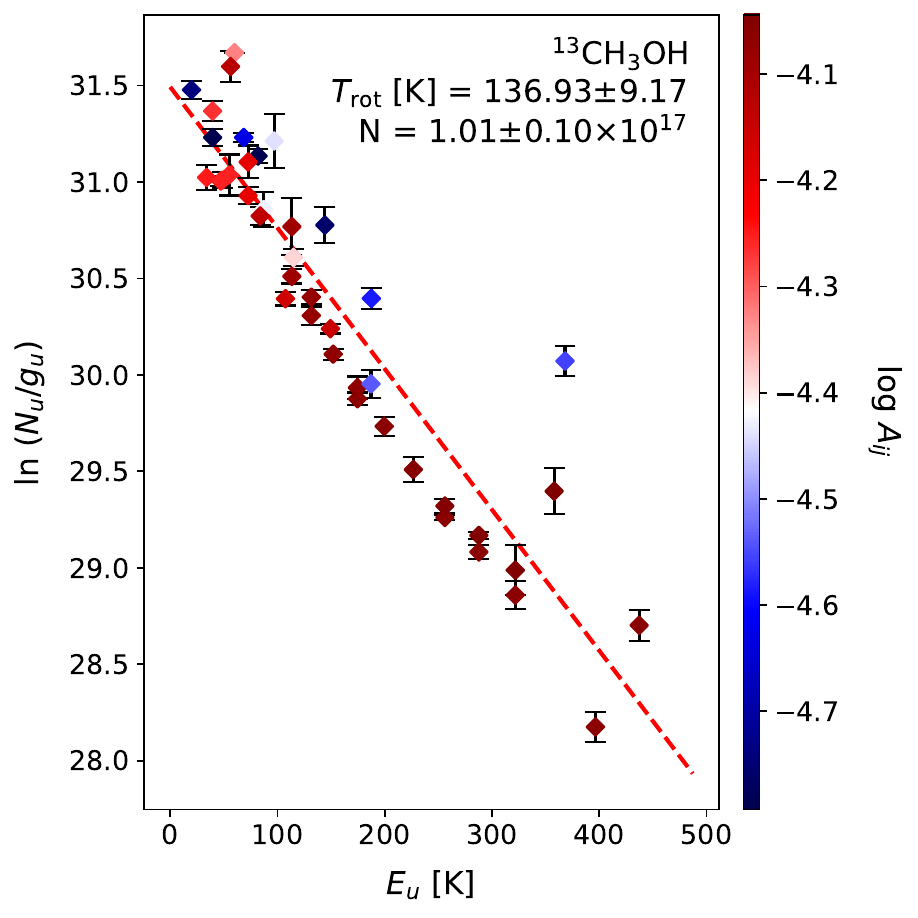}
    \includegraphics[trim=0.1cm 0cm 0cm 0cm, clip=true,width=0.325\textwidth]{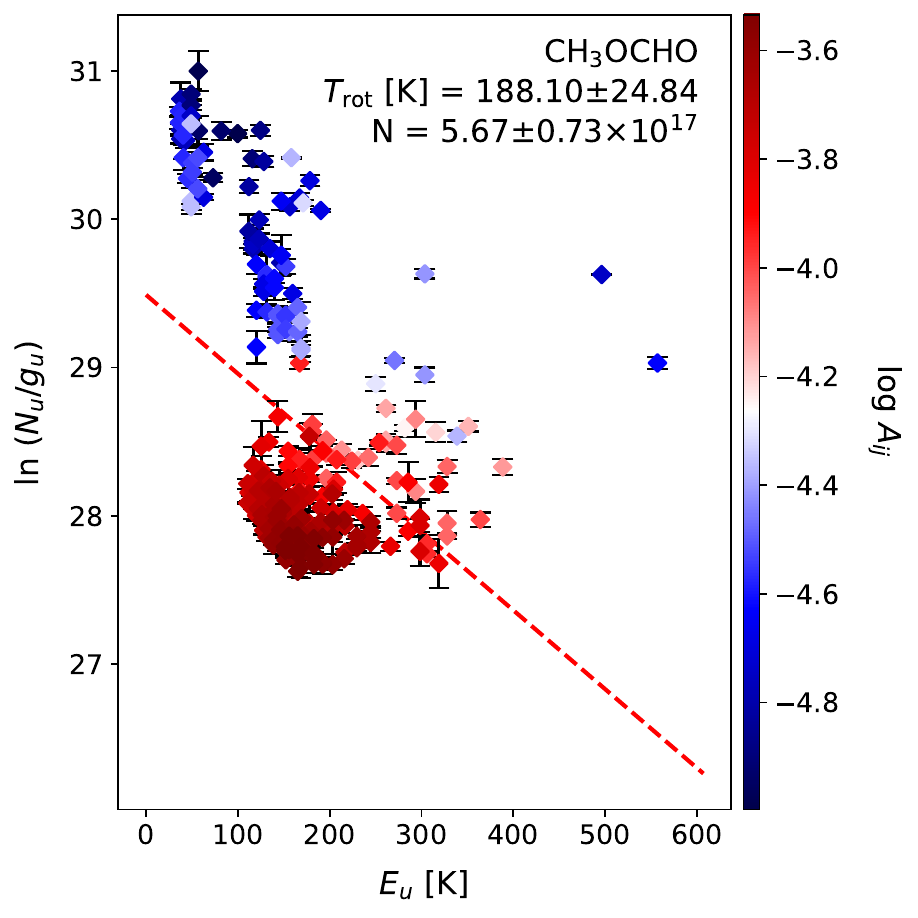}
    \includegraphics[trim=0.1cm 0cm 0cm 0cm, clip=true,width=0.325\textwidth]{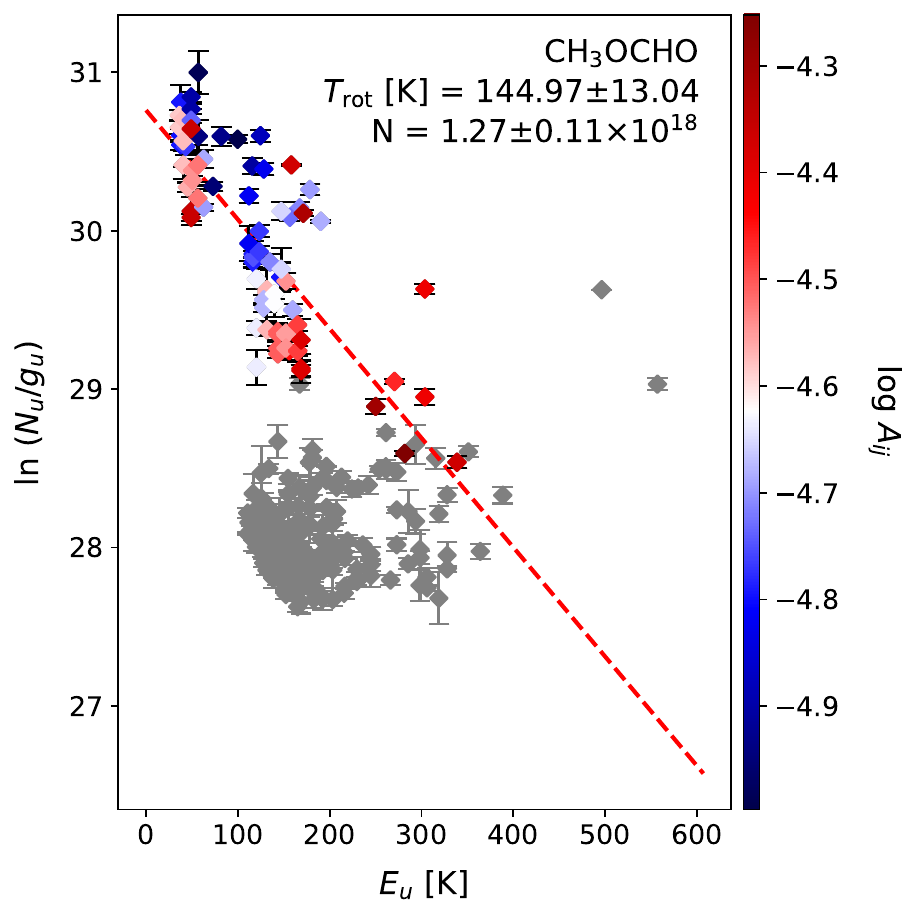}
    \caption{Example rotation diagrams for optically thin ($^{13}$CH$_{3}$OH, left panel) and thick (CH$_{3}$OCHO, center and right panels) molecules. On the right panel, the gray data points are lines with relatively high $A_{\rm ij}$, and those with color are lines with relatively low $A_{\rm ij}$. The red dashed line on the right panel is the result of a linear fit using only lines with low $A_{\rm ij}$ (colored points), while the gray dashed line shows the result when we use all lines (gray + colored points). This gray dashed line corresponds to the red dashed line in the middle panel, which shows the fitting to all data points.}
    \label{fig:rotation diagram examples}
\end{figure*}

The example plots for the rotation diagrams are shown in Figure \ref{fig:rotation diagram examples}. The rotation diagram shows two representative types. For molecules whose lines are optically thin, such as $^{13}$CH$_{3}$OH on the left panel of Figure \ref{fig:rotation diagram examples}, the linear fit works reasonably. However, for some COMs such as CH$_{3}$OCHO (middle, right panel), the lines are divided into two groups on the rotation diagram, resulting in an undesirable fit. The two groups have two distinct $A_{\rm ij}$ values. Lines with relatively higher $A_{\rm ij}$ show smaller column densities per degeneracy than those with relatively lower $A_{\rm ij}$ because they have higher optical depths (middle panel). Thus, we excluded the group with high $A_{\rm ij}$ and linear fitted to lines with low $A_{\rm ij}$, resulting in a higher column density by a few factors (right panel). 

\subsection{Results of rotation diagram analysis}
The molecular information could not be found for some molecules, such as C$_2$H$_3$CHO, so we did not fit their rotation diagrams. Molecules with less than 3 emission lines were also excluded. Finally, we reasonably fitted the rotation diagrams of 13 species.\par

The resulting column densities obtained from the rotation diagrams are listed in Table \ref{tb:rotation}. We used the column density of $^{13}$CH$_{3}$OH to calculate the abundances relative to CH$_{3}$OH of each molecule. The ratio between $^{13}$CH$_{3}$OH and CH$_{3}$OH was assumed as 1:24.4, as derived from iterative LTE line fitting results in Section \ref{sec:12C/13C}.\par

As described in Section \ref{sec:rotation diagram method}, we calculated the rotation temperatures and column densities using only lines with small $A_{\rm ij}$ for molecules with some of their molecular lines are attenuated. After fitting the rotation diagram, we checked if the lines with large $A_{\rm ij}$ are truly optically thick using \texttt{XCLASS}, with which we can calculate the line optical depths adopting the parameters derived from the rotation diagrams. As a result, there is a positive relation between the $A_{\rm ij}$ coefficient and the optical depth of the line. Therefore, using only small $A_{\rm ij}$ is almost equivalent to excluding lines with high optical depths.\par

\begin{figure}[t]
    \centering
    \includegraphics[trim=0.2cm 0cm 0.2cm 0cm, clip=true,width=0.48\textwidth]{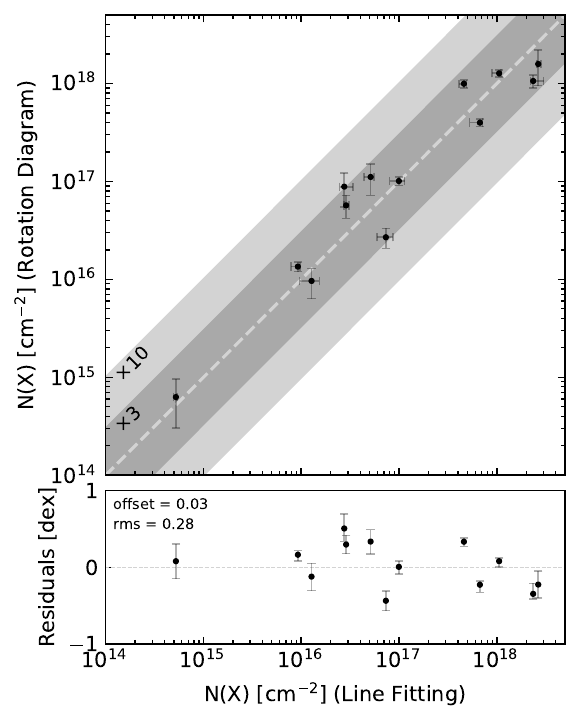}
    \caption{Comparison of column densities derived from the iterative LTE line fitting (Table \ref{tb:identified}) and from the rotation diagrams (Table \ref{tb:rotation}). The 1:1 reference line is depicted as a gray dashed line. Additionally, the gray solid lines, accompanied by dark and light gray shaded regions, represent three times and ten times the reference value.}
    \label{fig:comparison_XCLASS&RD}
\end{figure}

Lastly, in Figure \ref{fig:comparison_XCLASS&RD} we compared the column densities with the results from the iterative LTE line fitting. As described in the main paper, the temperature is fixed as 120 K in the line fitting method, although both the column density and rotation temperature are free parameters in rotation diagrams. The difference between the two methods has a scatter of about 0.28 dex, with a very little offset of 0.03 dex. Therefore, considering the column density range spans approximately four orders of magnitude, both methods can be regarded as consistent.

\startlongtable
    \begin{deluxetable*}{cccccccc}
    \tabletypesize{\scriptsize}
    \tablecaption{Rotation temperatures and column densities derived from rotation diagrams.\label{tb:rotation}}
    \tablehead{
    \colhead{Species} & \colhead{T$_{rot}$ (K)} & \colhead{N (cm$^{-2}$)} & \colhead{N(X)/N(CH$_{3}$OH)} & \colhead{Species} & \colhead{T$_{rot}$ (K)} & \colhead{N (cm$^{-2}$)} & \colhead{N(X)/N(CH$_{3}$OH)}}
    \startdata
    CH$_{3}$OH\tablenotemark{$\dag$} & 168.1(18.9) & $1.06(0.16)\times10^{18}$ & - & C$_{2}$H$_{5}$OH & 207.7(105) & $8.84(3.32)\times10^{16}$ & $3.6\times10^{-2}$\\
    CH$_{3}$OH v$_{12}$=1 & 137.0(15.7) & $1.58(0.63)\times10^{18}$ & - & s-C$_{2}$H$_{5}$CHO & 125.9(25.8) & $5.70(1.51)\times10^{16}$ & $2.3\times10^{-2}$\\
    $^{13}$CH$_{3}$OH & 136.9(9.17) & $1.01(0.10)\times10^{17}$ & $4.1\times10^{-2}$ & CH$_{3}$CCH & 93.6(25.0) & $9.64(3.27)\times10^{15}$ & $3.9\times10^{-3}$\\
    CH$_{3}$ $^{18}$OH & 165.1(49.0) & $2.69(0.62)\times10^{16}$ & $1.1\times10^{-2}$ & CH$_{3}$OCH$_{3}$ & 169.6(14.2) & $9.94(0.94)\times10^{17}$ & $4.0\times10^{-1}$\\
    CH$_{3}$OCHO\tablenotemark{$\dag$} & 145.0(13.0) & $1.27(0.11)\times10^{18}$ & $5.2\times10^{-1}$ & CH$_{3}$CN & 343.0(102) & $1.35(0.15)\times10^{16}$ & $5.5\times10^{-3}$\\
    c-C$_{2}$H$_{4}$O\tablenotemark{$\dag$} & 105.2(13.6) & $1.11(0.39)\times10^{17}$ & $4.5\times10^{-2}$ & $^{13}$CH$_{3}$CN & 179.0(127) & $6.29(3.27)\times10^{14}$ & $2.6\times10^{-4}$\\
    CH$_{3}$CHO\tablenotemark{$\dag$} & 134.9(12.3) & $3.98(0.33)\times10^{17}$ & $1.6\times10^{-1}$ & &  &  & \\
    \hline
    \enddata
    \tablenotetext{$\dag$}{Used lines with low Einstein-A coefficient (low optical depth).}
\end{deluxetable*}

\subsection{Rotation temperature for column density estimation} \label{sec:temperature variation}
Since Table \ref{tb:rotation} shows that the rotation diagrams result in higher temperatures than 120 K for some molecules, we investigated two higher rotation temperatures of 200 K and 300 K in the iterative line fitting process. For the investigation, we used different beam-filling factors (0.230, 0.154) to prevent optically thick lines from being calculated as optically thin due to the increased temperature and to keep the level of the saturated lines divided by the beam-filling factor similar to the rotation temperature. The $\chi^2$ value was used to find the best-fit temperature for observed spectra.\par

For the 200 K model, the $\chi^2$ for most species are found to be larger than those calculated at 120 K, while, for the 300 K case, all species show much larger $\chi^2$. The models with higher temperatures populate high-energy transition lines, which were not detected in the observed spectra.\par

There are five molecular species which fit better with 200K: CH$_{3}$OCHO v$_{18=1}$, CH$_{3}$CHO v$_{15=2}$, CH$_{3}$CN, CH$_{3}$CN v$_{8=1}$, and C$_{2}$H$_{5}$OH. These are mostly the vibrationally excited species that were previously found to have different column densities from the ground state species when 120 K was adopted (Section \ref{sec:column density result}). The column densities of these species increase slightly (e.g., from 6.9$\times$10$^{17}$ cm$^{-2}$ to 8.9$\times$10$^{17}$ cm$^{-2}$ for CH$_{3}$OCHO v$_{18=1}$; from 9.2$\times$10$^{15}$ cm$^{-2}$ to 1.1$\times$10$^{16}$ cm$^{-2}$ for CH$_{3}$CN), except for CH$_{3}$CN v$_{8=1}$. In fact, the column density of CH$_{3}$CN v$_{8=1}$ reduces to 1.1$\times$10$^{16}$ cm$^{-2}$ from 2.3$\times$10$^{16}$ cm$^{-2}$. In general, the column densities of the ground state and excited state species, which show some difference at the fixed excitation temperature of 120 K in Section \ref{sec:column density result}, become more consistent when 200 K is adopted for the excited species.\par 

Although the spectra of several molecules in excited vibrational states are fitted better with a higher temperature of 200 K as presented above, we opt to use a fixed temperature of 120 K to fit all molecular spectra since we extracted spectra over the entire disk and the average dust temperature over the disk is 120 K \citep{jelee19}. Above all, the spectra of most species are best fitted by 120 K. In addition, even for the five species mentioned above, the column densities fitted with 120 K and 200 K differ only by less than a factor of 2. As a result, our conclusion is not affected significantly by our choice of rotation temperature.

\subsection{Spatial distribution of COMs}
While we estimated single column density and rotation temperature per molecule using the averaged spectra extracted from the PC1-filtering method, the same procedure can be easily applied in each pixel to map the physical parameters using the rotation diagram fitting if the lines are strong enough over multiple pixels. However, the line fitting analysis on pixel-by-pixel requires much more computational time.\par 

From the original $200\times200$ pixels map with a pixel size of 0.02$\arcsec$, a smaller map of $15\times15$ pixels with the $4\times4$ binning was made to reduce noise and focus on the central region with COMs emission. In this map, the pixel size is comparable with the beam size. The source has a disk rotation, which has to be corrected to identify each emission line before constructing the rotation diagram. For the velocity correction, we used the same isolated, strong lines as used for the PCA analysis \citep{yun2023pca}.\par 

\begin{figure}
    \includegraphics[trim=0cm 0.2cm 0cm 0.2cm, clip=true,width=0.49\textwidth]{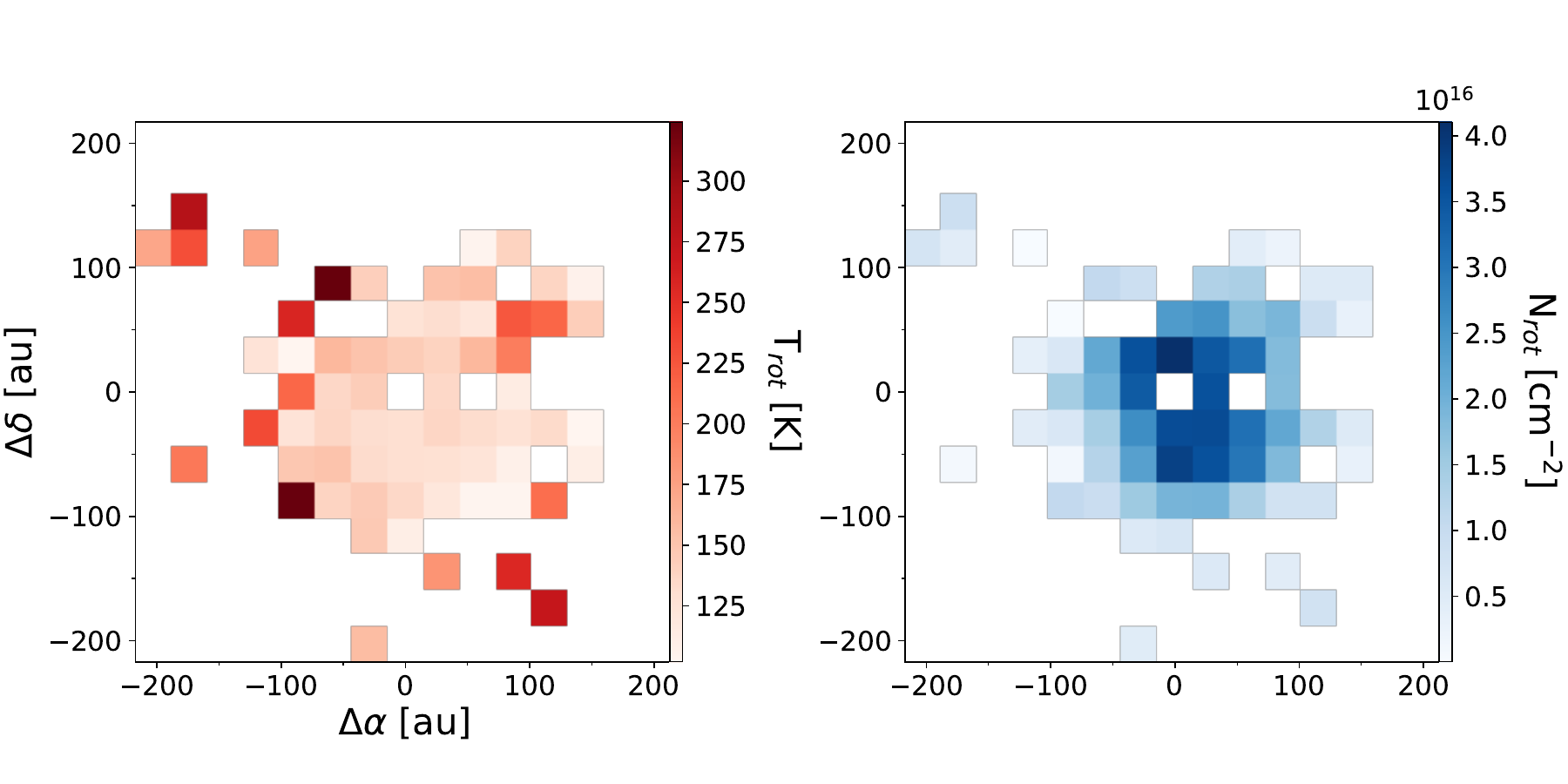}
    \includegraphics[trim=0cm 0.2cm 0cm 0.2cm, clip=true,width=0.49\textwidth]{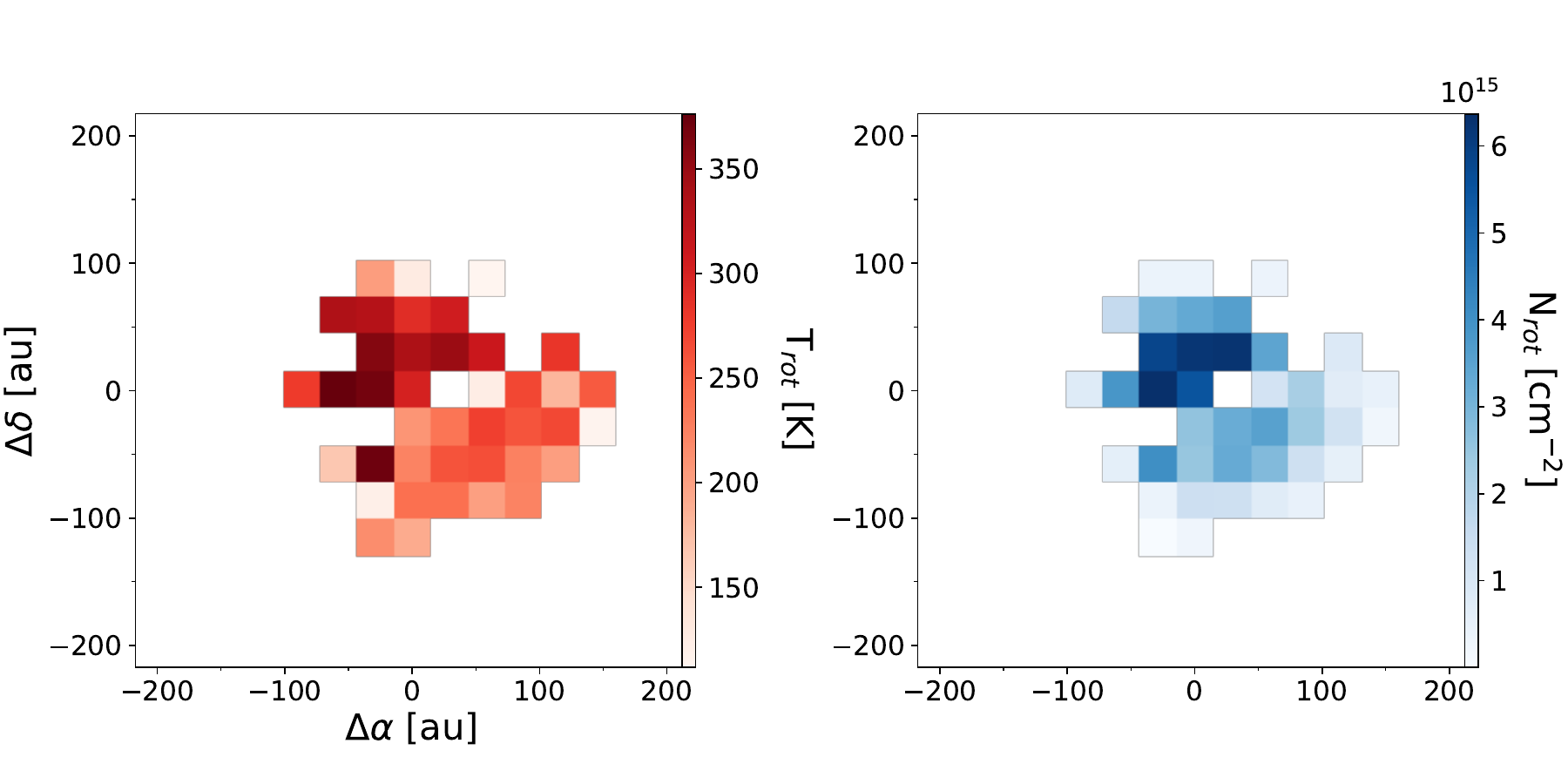}
    \caption{Rotation temperature (left) and column density (right) maps of $^{13}$CH$_3$OH (top) and CH$_3$CN (bottom). Only pixels with fitted temperatures between 100 K and 400 K are drawn.}
\label{fig:rot_map}
\end{figure}

The column density and rotation temperature maps of $^{13}$CH$_{3}$OH and CH$_{3}$CN are plotted in Figure \ref{fig:rot_map}. The maps for $^{13}$CH$_{3}$OH show a consistent rotation temperature of $\sim$120 K throughout the map. The column density distribution of $^{13}$CH$_{3}$OH shows no clear asymmetry. On the other hand, the column density and rotation temperature of CH$_{3}$CN are higher in the northern part. The temperature is much higher than 150 K in most pixels, and it is even higher than 300 K in some of the northern region. This might suggest that the origin of CH$_3$CN could be distinct from that of other oxygen-bearing COMs. This is also consistent with the results from the averaged-spectra in Appendix \ref{sec:temperature variation}, where 200 K fits better the CH$_{3}$CN spectra than 120 K. 

\section{Line optical depth effects seen with spectra}\label{sec:effect of line optical depth seen with spectra}

\begin{figure*}[t]
    \centering
    \includegraphics[trim=0.2cm 0cm 0.2cm 0cm, clip=true,width=1\textwidth]{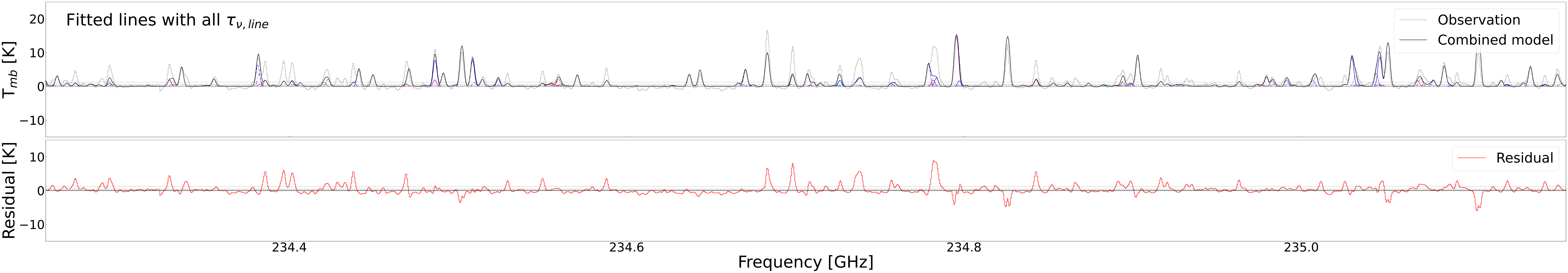}
    \includegraphics[trim=0.2cm 0cm 0.2cm 0cm, clip=true,width=1\textwidth]{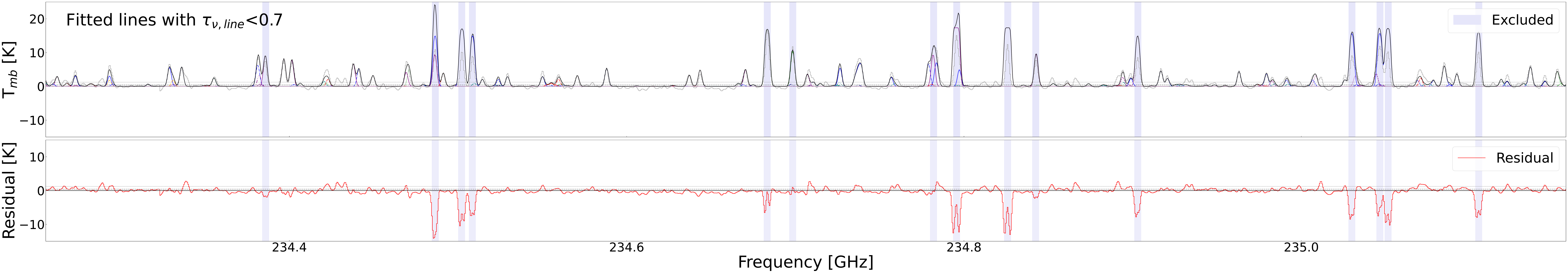}
    \caption{LTE line-fitted model spectra when using isolated molecular lines of all line optical depths (upper panel of this figure, and x-axis of Figure \ref{fig:comparison_different_tau}), and when using isolated lines of $\tau_{\rm \nu,line}<0.7$ (lower panel of this figure, and y-axis of Figure \ref{fig:comparison_different_tau}). Both panels show the same spectral window. Black solid lines represent the combined spectra of the model spectra for each species, and the individual model spectra are shown as various line colors and styles. The red solid lines indicate residual between the observed data and the model. The gray dashed/dotted horizontal lines represent the 1$\sigma$/3$\sigma$ levels of the observed spectrum. Also, the blue shaded areas on the lower panel indicate lines excluded during the LTE line fitting because their modeled optical depths exceeded 0.7. This frequency region is the same as the frequency region in the lower panel of Figure \ref{fig:spectrum_before}.}
    \label{fig:fitted_spectra_comparison}
\end{figure*}

Figure \ref{fig:fitted_spectra_comparison} illustrates how significantly the spectral fitting results can differ depending on the constraint of line optical depths. The model spectra in the lower panel fit well most of the observed spectral lines, particularly the weak lines, whereas the model spectra in the upper panel fit only the few strongest lines. Likewise, the iterative fitting process only with optically thin lines enables us to identify lines that might have been missed otherwise and, thus, derive the COMs column densities more precisely.\par 

Ideally, the model spectra fitting the weak lines must also fit the optically thick strong lines. However, the intensities of some optically thick model lines exceed those of corresponding observed lines (i.e., the lines excluded during the line fitting), as seen in the lower panel and its residual plot. This may be attributed to different beam-filling factors from what we adopted for the model spectra. Therefore, we test the effect of different beam-filling factors on the derived column densities and summarize the results in Appendix \ref{sec:effect of parameters}.\par

\section{Beam-filling factor for column density estimation}\label{sec:effect of parameters}
Here, we investigate the beam-filling factor to fit both optically thick/saturated and thin lines for one exemplary species, methyl formate (CH$_{3}$OCHO). The beam-filling factor scales the whole spectrum, determining the maximum intensities of the saturated lines along with the rotation temperature. The rotation temperature also controls the line ratio in the LTE condition; as presented in appendix \ref{sec:rotation diagram}, 120 K fits well the averaged spectra of most species.\par

Firstly, we test the beam-filling factor using a rotation diagram. The observational lines used in the rotation diagram analysis, where the lines are assumed to be optically thin and under the LTE condition, will be weaker than the expected line intensity by a factor of $C=\tau_{\rm \nu, line}/(1-e^{-\tau_{\rm \nu, line}})$. In reality, since the optical depth for each emission line is different, this causes uncertainty in the temperature and the column density derived from the rotation diagram. This effect is clearly seen in the rotation diagram of methyl formate in Figure \ref{fig:rotation diagram examples}. The data points are divided into two groups because the lines with high optical depth (red symbols) have been attenuated by self-absorption.\par

We can calculate the line optical depth with \texttt{XCLASS} adopting the temperature and column density derived from the rotation diagram (Table \ref{tb:rotation}) and recalculate the intensities of each line after correcting for the optical depth effect. However, correction by the calculated optical depths did not merge the two groups into a linear line. Many lines with high A$_{ij}$ had optical depth around $\tau \sim 3.5$, which corresponds to a factor of $\sim$3.5 in the column densities in the rotation diagram. The column density must be higher than the current value (1.27$\times$10$^{18}$ cm$^{-2}$) by a factor of 10/6 to raise the group of optically thick lines (red symbols) and merge with the group of optically thin lines (blue symbols).\par

However, the higher column density also affects other lines. Here, we can play with the beam-filling factor. As mentioned above, the beam-filling factor is set to 0.384, but a value of 0.23 (0.384$\times$0.6) resulted in a better linear fit for the rotation diagram of CH$_3$OCHO.\par

In the intensity maps of some CH$_3$OCHO lines (Paper I), the emission distributes along a ring structure with the outer radius of $\sim$0.2$\arcsec$. The inner radius of the ring structure is 0.1$\arcsec$, which is caused by the optically thick dust continuum in the inner region ($<$ 0.1$\arcsec$) \citep{cieza_2016}. Since the spectrum was extracted from a COMs-rich region with a radius of about 0.3$\arcsec$, following the definition in \citet{Moller17}, the beam-filling factor of the CH$_3$OCHO emission is around 1/4, similar to 0.23.\par 

We tested line fitting by adopting the beam-filling factor of 0.23, although the beam-filling factor could vary with transitions even in a molecular species. The rotation temperature of 120 K and beam-filling factor of 0.23 match the upper bound of saturated lines of methyl formate, which is about 28 K after the dust attenuation correction. Thus, the line fitting was done with one free parameter (column density) only to isolated, optically thin lines of methyl formate. As a result, the model spectra well fitted both optically thick, saturated lines and weak lines. The obtained column density is 2.1$\times$10$^{18}$ cm$^{-2}$, which is about 1.9 times higher than the value obtained in the line fitting method (Table \ref{tb:identified}), and 1.7 times higher than the rotation diagram method (Table \ref{tb:rotation}). The optical depths of optically thick lines in this model increase to 6--7, as intended.\par

The emission size of most molecules is likely similar to or smaller than the size of CH$_3$OH emission, which is used as a standard. Therefore, the accurate beam-filling factors of less abundant molecules can increase the column densities by a factor of $\sim2-3$, so the current values can be considered lower limits. However, the beam-filling factor varies with both molecules and transitions. If we perform line fitting with beam-filling factor and column density as free parameters, the parameters do not converge well.\par
Therefore, we leave this as a caveat of the current analysis but note that the column density ratios such as isotope ratios would not change significantly because the increase of column density by a smaller beam-filling factor would not vary much from molecule to molecule.

\section{Comparison with ALMA/PILS survey} \label{sec:long table}
In this Appendix, we provide a table that was discussed in Section \ref{sec:comparison against hot corino and comet}.

\startlongtable
\begin{deluxetable*}{ccccccc}
    \tabletypesize{\scriptsize}
    \tablecaption{Comparison of the observed COMs and their abundances in V883 Ori and IRAS 16293-2422B.\label{tb:comparisons}}
    \tablehead{
    \colhead{No.} & \colhead{Species}& \multicolumn{2}{c}{N (cm$^{-2}$)} & \multicolumn{2}{c}{N(X)/N(CH$_{3}$OH)} & \colhead{Reference\tablenotemark{a}}\\\colhead{}&\colhead{}&\colhead{V883 Ori\tablenotemark{b}}&  \colhead{IRAS 16293-2422B}& \colhead{V883 Ori\tablenotemark{c}}&  \colhead{IRAS 16293-2422B} & \colhead{}}
    \startdata
    \multicolumn{7}{c}{Methanol}\\
	1 & CH$_{3}$OH & $2.3\times10^{18}$ & $1.0\times10^{19}$ & 1.0 & 1.0 & 11 \\
	 & CH$_{3}$OH v$_{12}$=1 & $2.6\times10^{18}$ & --\tablenotemark{d} & $1.1\times10^{+0}$ & -- & -- \\
	 & CH$_{3}$OH v$_{12}$=2 & $2.1\times10^{18}$ & -- & $9.1\times10^{-1}$ & -- & -- \\
	2 & $^{13}$CH$_{3}$OH & $9.9\times10^{16}$ & -- & $4.3\times10^{-2}$ & -- & 1 \\
	3 & CH$_{2}$DOH & $7.5\times10^{16}$ & $7.1\times10^{17}$ & $3.3\times10^{-2}$ & $7.1\times10^{-2}$ & 11 \\
	4 & CHD$_{2}$OH & $< 1.2\times10^{16}$\tablenotemark{e} & $1.8\times10^{17}$ & $< 5.2\times10^{-3}$ & $1.8\times10^{-2}$ & 19 \\
	5 & CH$_{3}$OD & -- & $1.8\times10^{17}$ & -- & $1.8\times10^{-2}$ & 11 \\
	6 & CD$_{3}$OH & $< 6.2\times10^{15}$ & $3.1\times10^{16}$ & $< 2.7\times10^{-3}$ & $3.1\times10^{-3}$ & 18 \\
	7 & CD$_{3}$OD & -- & $< 2.0\times10^{15}$ & -- & $< 2.0\times10^{-4}$ & 20 \\
	8 & CH$_{3}$$^{18}$OH & $7.3\times10^{16}$ & $1.9\times10^{16}$ & $3.2\times10^{-2}$ & $1.9\times10^{-3}$ & 3 \\
     \multicolumn{7}{c}{Methyl formate}\\
	9 & CH$_{3}$OCHO & $1.1\times10^{18}$ & $2.6\times10^{17}$ & $4.8\times10^{-1}$ & $2.6\times10^{-2}$ & 11 \\
      & CH$_{3}$OCHO v$_{18}$=1 & $6.9\times10^{17}$ & -- & $3.0\times10^{-1}$ & -- & -- \\
	10 & CH$_{3}$OCDO & -- & $1.5\times10^{16}$ & -- & $1.5\times10^{-3}$ & 11 \\
	11 & CH$_{2}$DOCHO & -- & $4.8\times10^{16}$ & -- & $4.8\times10^{-3}$ & 11 \\
	12 & CH$_{3}$O$^{13}$CHO & $7.4\times10^{16}$ & $< 6.3\times10^{15}$ & $3.2\times10^{-2}$ & $< 6.3\times10^{-4}$ & 11 \\
	13 & CHD$_{2}$OCHO & -- & $1.1\times10^{16}$ & -- & $1.1\times10^{-3}$ & 12 \\
     \multicolumn{7}{c}{Ethanol}\\
	14 & C$_{2}$H$_{5}$OH & $2.8\times10^{16}$ & $2.3\times10^{17}$ & $1.2\times10^{-2}$ & $2.3\times10^{-2}$ & 11 \\
	15 & a-a-CH$_{2}$DCH$_{2}$OH & $< 6.5\times10^{15}$ & $2.7\times10^{16}$ & $< 2.8\times10^{-3}$ & $2.7\times10^{-3}$ & 11 \\
	16 & a-s-CH$_{2}$DCH$_{2}$OH & $< 6.5\times10^{15}$ & $1.3\times10^{16}$ & $< 2.8\times10^{-3}$ & $1.3\times10^{-3}$ & 11 \\
	17 & a-CH$_{3}$CHDOH & $< 6.4\times10^{15}$ & $2.3\times10^{16}$ & $< 2.8\times10^{-3}$ & $2.3\times10^{-3}$ & 11 \\
	18 & a-CH$_{3}$CH$_{2}$OD & $< 6.4\times10^{15}$ & $< 1.1\times10^{16}$ & $< 2.8\times10^{-3}$ & $< 1.1\times10^{-3}$ & 11 \\
	19 & a-CH$_{3}$$^{13}$CH$_{2}$OH & $< 6.1\times10^{15}$ & $< 9.1\times10^{15}$ & $< 2.7\times10^{-3}$ & $< 9.1\times10^{-4}$ & 11 \\
	20 & a-$^{13}$CH$_{3}$CH$_{2}$OH & $< 6.2\times10^{15}$ & $< 9.1\times10^{15}$ & $< 2.7\times10^{-3}$ & $< 9.1\times10^{-4}$ & 11 \\
     \multicolumn{7}{c}{Acetaldehyde}\\
	21 & CH$_{3}$CHO & $6.7\times10^{17}$ & $1.2\times10^{17}$ & $2.9\times10^{-1}$ & $1.2\times10^{-2}$ & 4 \\
       & CH$_{3}$CHO v$_{15}$=1 & $5.7\times10^{17}$ & -- & $2.5\times10^{-1}$ & -- & -- \\
       & CH$_{3}$CHO v$_{15}$=2 & $4.0\times10^{17}$ & -- & $1.7\times10^{-1}$ & -- & -- \\
	22 & CH$_{3}$CDO & $9.8\times10^{15}$ & $7.4\times10^{15}$ & $4.3\times10^{-3}$ & $7.4\times10^{-4}$ & 11 \\
	23 & $^{13}$CH$_{3}$CHO & $1.8\times10^{16}$ & $1.8\times10^{15}$ & $7.8\times10^{-3}$ & $1.8\times10^{-4}$ & 11 \\
       & $^{13}$CH$_{3}$CHO v$_{15}$=1 & $1.8\times10^{16}$ & -- & $7.8\times10^{-3}$ & -- & -- \\
	24 & CH$_{3}$$^{13}$CHO & $1.7\times10^{16}$ & $1.8\times10^{15}$ & $7.4\times10^{-3}$ & $1.8\times10^{-4}$ & 11 \\
       & CH$_{3}$$^{13}$CHO v$_{15}$=1 & $2.0\times10^{16}$ & -- & $8.7\times10^{-3}$ & -- & -- \\
	25 & CH$_{2}$DCHO & $2.1\times10^{16}$ & $6.2\times10^{15}$ & $9.1\times10^{-3}$ & $6.2\times10^{-4}$ & 16 \\
     \multicolumn{7}{c}{Acetone}\\
	26 & CH$_{3}$COCH$_{3}$ & $4.4\times10^{16}$ & $1.7\times10^{16}$ & $1.9\times10^{-2}$ & $1.7\times10^{-3}$ & 4 \\
     \multicolumn{7}{c}{Ethylene glycol}\\
	27 & aGg'-(CH$_{2}$OH)$_2$ & $< 5.5\times10^{15}$ & $5.2\times10^{16}$ & $< 2.4\times10^{-3}$ & $5.2\times10^{-3}$ & 3 \\
	28 & gGg'-(CH$_{2}$OH)$_2$ & $< 9.8\times10^{15}$ & $4.7\times10^{16}$ & $< 4.3\times10^{-3}$ & $4.7\times10^{-3}$ & 3 \\
     \multicolumn{7}{c}{Dimethyl ether}\\
	29 & CH$_{3}$OCH$_{3}$ & $4.6\times10^{17}$ & $2.4\times10^{17}$ & $2.0\times10^{-1}$ & $2.4\times10^{-2}$ & 11 \\
	30 & $^{13}$CH$_{3}$OCH$_{3}$ & -- & $1.4\times10^{16}$ & -- & $1.4\times10^{-3}$ & 11 \\
	31 & asym-CH$_{2}$DOCH$_{3}$ & -- & $4.1\times10^{16}$ & -- & $4.1\times10^{-3}$ & 11 \\
	32 & sym-CH$_{2}$DOCH$_{3}$ & -- & $1.2\times10^{16}$ & -- & $1.2\times10^{-3}$ & 11 \\
     \multicolumn{7}{c}{Glycolaldehyde}\\
	33 & CH$_{2}$OHCHO & $< 1.7\times10^{15}$ & $3.2\times10^{16}$ & $< 7.4\times10^{-4}$ & $3.2\times10^{-3}$ & 3 \\
	34 & CHDOHCHO & -- & $3.3\times10^{15}$ & -- & $3.3\times10^{-4}$ & 3 \\
	35 & CH$_{2}$ODCHO & -- & $1.5\times10^{15}$ & -- & $1.5\times10^{-4}$ & 3 \\
	36 & CH$_{2}$OHCDO & -- & $1.6\times10^{15}$ & -- & $1.6\times10^{-4}$ & 3 \\
	37 & $^{13}$CH$_{2}$OHCHO & -- & $1.2\times10^{15}$ & -- & $1.2\times10^{-4}$ & 3 \\
	38 & CH$_{2}$OH$^{13}$CHO & -- & $1.2\times10^{15}$ & -- & $1.2\times10^{-4}$ & 3 \\
     \multicolumn{7}{c}{Acetic acid}\\
	39 & CH$_{3}$COOH & $< 1.3\times10^{16}$ & $2.8\times10^{15}$ & $< 5.7\times10^{-3}$ & $2.8\times10^{-4}$ & 3 \\
     \multicolumn{7}{c}{Propanal}\\
	40 & C$_{2}$H$_{5}$CHO & $2.9\times10^{16}$ & $2.2\times10^{15}$ & $1.3\times10^{-2}$ & $2.2\times10^{-4}$ & 4 \\
     \multicolumn{7}{c}{Ethylene oxide}\\
	41 & c-C$_{2}$H$_{4}$O & $5.1\times10^{16}$ & $5.4\times10^{15}$ & $2.2\times10^{-2}$ & $5.4\times10^{-4}$ & 4 \\
	42 & c-C$^{13}$CH$_{4}$O & $5.3\times10^{15}$ & -- & $2.3\times10^{-3}$ & -- & -- \\
     \multicolumn{7}{c}{Trans-ethyl methyl ether}\\
	43 & t-C$_{2}$H$_{5}$OCH$_{3}$ & -- & $1.8\times10^{16}$ & -- & $1.8\times10^{-3}$ & 16 \\
     \multicolumn{7}{c}{Methoxymethanol}\\
	44 & CH$_{3}$OCH$_{2}$OH & $< 3.0\times10^{17}$ & $1.4\times10^{17}$ & $< 1.3\times10^{-1}$ & $1.4\times10^{-2}$ & 16 \\
     \multicolumn{7}{c}{2-Propenal}\\
	45 & C$_{2}$H$_{3}$CHO & $7.7\times10^{15}$ & $3.4\times10^{14}$ & $3.3\times10^{-3}$ & $3.4\times10^{-5}$ & 17 \\
     \multicolumn{7}{c}{Isopropyl alcohol}\\
	46 & n-C$_{3}$H$_{7}$OH & $< 1.1\times10^{16}$ & $< 3.0\times10^{15}$ & $< 4.8\times10^{-3}$ & $< 3.0\times10^{-4}$ & 17 \\
	47 & i-C$_{3}$H$_{7}$OH\tablenotemark{f} & $< 4.3\times10^{16}$ & $< 3.0\times10^{15}$ & $< 1.9\times10^{-2}$ & $< 3.0\times10^{-4}$ & 17 \\
     \multicolumn{7}{c}{2-Propynal}\\
	48 & HCCCHO & $< 2.2\times10^{15}$ & $< 5.0\times10^{14}$ & $< 9.6\times10^{-4}$ & $< 5.0\times10^{-5}$ & 17 \\
     \multicolumn{7}{c}{Glyoxal}\\
	49 & cis-HC(O)CHO & -- & $< 5.0\times10^{13}$ & -- & $< 5.0\times10^{-6}$ & 17 \\
     \multicolumn{7}{c}{Propyne}\\
	50 & CH$_{3}$CCH & $1.3\times10^{16}$ & $6.8\times10^{15}$ & $5.7\times10^{-3}$ & $6.8\times10^{-4}$ & 15 \\
	51 & CH$_{3}^{13}$CCH & $< 4.8\times10^{15}$ & $< 2.4\times10^{14}$ & $< 2.1\times10^{-3}$ & $< 2.4\times10^{-5}$ & 15 \\
	52 & $^{13}$CH$_{3}$CCH & $< 4.9\times10^{15}$ & $< 2.5\times10^{14}$ & $< 2.1\times10^{-3}$ & $< 2.5\times10^{-5}$ & 15 \\
	53 & CH$_{3}$C$^{13}$CH & $< 5.0\times10^{15}$ & $< 2.5\times10^{14}$ & $< 2.2\times10^{-3}$ & $< 2.5\times10^{-5}$ & 15 \\
	54 & CH$_{3}$CCD & $< 5.3\times10^{15}$ & $< 7.9\times10^{14}$ & $< 2.3\times10^{-3}$ & $< 7.9\times10^{-5}$ & 15 \\
	55 & CH$_{2}$DCCH & $< 5.3\times10^{15}$ & $< 2.6\times10^{14}$ & $< 2.3\times10^{-3}$ & $< 2.6\times10^{-5}$ & 15 \\
     \multicolumn{7}{c}{Propylene}\\
	56 & C$_{3}$H$_{6}$ & -- & $4.2\times10^{16}$ & -- & $4.2\times10^{-3}$ & 17 \\
     \multicolumn{7}{c}{Propane}\\
	57 & C$_{3}$H$_{8}$ & $< 1.9\times10^{18}$ & $< 8.0\times10^{16}$ & $< 8.3\times10^{-1}$ & $< 8.0\times10^{-3}$ & 17 \\
     \multicolumn{7}{c}{Methyl cyanide}\\
	58 & CH$_{3}$CN & $9.2\times10^{15}$ & $4.0\times10^{16}$ & $4.0\times10^{-3}$ & $4.0\times10^{-3}$ & 7 \\
       & CH$_{3}$CN v$_{8}$=1 & $2.3\times10^{16}$ & $4.0\times10^{16}$ & $1.0\times10^{-2}$ & $4.0\times10^{-3}$ & 7 \\
	59 & $^{13}$CH$_{3}$CN & $5.2\times10^{14}$ & $6.0\times10^{14}$ & $2.3\times10^{-4}$ & $6.0\times10^{-5}$ & 7 \\
	60 & CH$_{3}$$^{13}$CN & $4.9\times10^{14}$ & $5.0\times10^{14}$ & $2.1\times10^{-4}$ & $5.0\times10^{-5}$ & 7 \\
	61 & CH$_{3}$C$^{15}$N & $< 1.9\times10^{14}$ & $1.6\times10^{14}$ & $< 8.3\times10^{-5}$ & $1.6\times10^{-5}$ & 7 \\
	62 & CH$_{2}$DCN & $1.2\times10^{15}$ & $1.4\times10^{15}$ & $5.2\times10^{-4}$ & $1.4\times10^{-4}$ & 7 \\
	63 & CHD$_{2}$CN & $< 3.2\times10^{14}$ & $2.0\times10^{14}$ & $< 1.4\times10^{-4}$ & $2.0\times10^{-5}$ & 7 \\
     \multicolumn{7}{c}{Methyl isocyanide}\\	
	64 & CH$_{3}$NC & $< 1.8\times10^{14}$ & $2.0\times10^{14}$ & $< 7.8\times10^{-5}$ & $2.0\times10^{-5}$ & 8 \\
     \multicolumn{7}{c}{Formamide}\\
	65 & NH$_{2}$CHO & $< 5.0\times10^{14}$ & $9.5\times10^{15}$ & $< 2.2\times10^{-4}$ & $9.5\times10^{-4}$ & 2 \\
	66 & NH$_{2}$CDO & $< 5.6\times10^{14}$ & $2.1\times10^{14}$ & $< 2.4\times10^{-4}$ & $2.1\times10^{-5}$ & 2 \\
	67 & cis-NHDCHO & $< 5.4\times10^{14}$ & $2.1\times10^{14}$ & $< 2.3\times10^{-4}$ & $2.1\times10^{-5}$ & 2 \\
	68 & trans-NHDCHO & $< 5.2\times10^{14}$ & $1.8\times10^{14}$ & $< 2.3\times10^{-4}$ & $1.8\times10^{-5}$ & 2 \\
	69 & NH$_{2}$$^{13}$CHO & $< 4.9\times10^{14}$ & $1.5\times10^{14}$ & $< 2.1\times10^{-4}$ & $1.5\times10^{-5}$ & 2 \\
	70 & $^{15}$NH$_{2}$CHO & $< 5.0\times10^{14}$ & $< 1.0\times10^{14}$ & $< 2.2\times10^{-4}$ & $< 1.0\times10^{-5}$ & 2 \\
	71 & NH$_{2}$CH$^{18}$O & $< 5.2\times10^{14}$ & $< 0.8\times10^{14}$ & $< 2.3\times10^{-4}$ & $< 8.0\times10^{-6}$ & 2 \\
     \multicolumn{7}{c}{Glycine}\\
	72 & NH$_{2}$CH$_{2}$COOH & $< 2.2\times10^{16}$ & $< 9.2\times10^{14}$ & $< 9.6\times10^{-3}$ & $< 9.2\times10^{-5}$ & 14 \\
     \multicolumn{7}{c}{Acetamide}\\
	73 & CH$_{3}$C(O)NH$_{2}$ & -- & $< 9.0\times10^{14}$ & -- & $< 9.0\times10^{-5}$ & 10 \\
     \multicolumn{7}{c}{Methyl isocyanate}\\
	74 & CH$_{3}$NCO & $< 2.2\times10^{16}$ & $4.0\times10^{15}$ & $< 9.6\times10^{-3}$ & $4.0\times10^{-4}$ & 5 \\
	75 & CH$_{3}$CNO & $< 2.9\times10^{14}$ & $< 5.0\times10^{13}$ & $< 1.3\times10^{-4}$ & $< 5.0\times10^{-6}$ & 5 \\
	76 & CH$_{3}$OCN & $< 1.0\times10^{15}$ & $< 5.0\times10^{14}$ & $< 4.3\times10^{-4}$ & $< 5.0\times10^{-5}$ & 5 \\
     \multicolumn{7}{c}{Methylamine}\\
	77 & CH$_{3}$NH$_{2}$ & $< 1.3\times10^{16}$ & $< 5.3\times10^{14}$ & $< 5.7\times10^{-3}$ & $< 5.3\times10^{-5}$ & 9 \\
     \multicolumn{7}{c}{Vinyl cyanide}\\
	78 & C$_{2}$H$_{3}$CN & $< 9.5\times10^{14}$ & $7.4\times10^{14}$ & $< 4.1\times10^{-4}$ & $7.4\times10^{-5}$ & 7 \\
     \multicolumn{7}{c}{Ethyl cyanide}\\ 
	79 & C$_{2}$H$_{5}$CN & $< 1.1\times10^{15}$ & $3.6\times10^{15}$ & $< 4.8\times10^{-4}$ & $3.6\times10^{-4}$ & 7 \\
     \multicolumn{7}{c}{Methyl mercaptan}\\
	80 & CH$_{3}$SH & -- & $4.8\times10^{15}$ & -- & $4.8\times10^{-4}$ & 6 \\
	81 & CH$_{3}$SD & $< 5.4\times10^{15}$ & $< 8.8\times10^{14}$ & $< 2.3\times10^{-3}$ & $< 8.8\times10^{-5}$ & 13 \\
    \hline
    \multicolumn{7}{p{14cm}}{References. (1) \citealt{PILS0}; (2) \citealt{PILS1}; (3) \citealt{PILS2}; (4) \citealt{PILS3}; (5) \citealt{PILS4}; (6) \citealt{PILS8}; (7) \citealt{PILS10}; (8) \citealt{PILS11}; (9) \citealt{PILS12}; (10) \citealt{PILS13}; (11) \citealt{PILS14}; (12) \citealt{PILS15}; (13) \citealt{PILS16}; (14) \citealt{PILS19}; (15) \citealt{PILS20}; (16) \citealt{PILS21}; (17) \citealt{PILS22}; (18) \citealt{PILS23}; (19) \citealt{PILS24}; (20) \citealt{Ilyushin_2023}.} 
    \enddata
    \tablenotetext{a}{References for IRAS 16293-2422B.}
    \tablenotetext{b}{Used column densities derived using LTE line fitting method (Table \ref{tb:identified}).}
    \tablenotetext{c}{Since only optically thin lines were used in the line fitting, we adopted the column density value for CH$_{3}$OH as it is, and then divided it directly to calculate the CH$_{3}$OH relative abundance of the other molecules.}
    \tablenotetext{d}{The -- represents that there is no information on whether this species has been detected or not.}
    \tablenotetext{e}{3$\sigma$ upper limits for the non-detected molecules.}
    \tablenotetext{f}{N(g-i-C$_3$H$_7$OH) $+$ N(a-i-C$_3$H$_7$OH)}
\end{deluxetable*}

\end{document}